%% file: karaenkeSchifferWaldherr.tex
\documentclass[a4paper, 12pt, abstracton, notitlepage, oneside]{scrartcl}
\makeatletter

\usepackage{lmodern}

\usepackage[T1]{fontenc}
\usepackage[utf8]{inputenc}
\usepackage[onehalfspacing]{setspace}
\usepackage[headsepline]{scrlayer-scrpage}
\clearscrheadfoot 
\ohead{\pagemark}
\usepackage{layout}
\usepackage[english]{babel}
\usepackage[left=2.54cm,right=2.54cm,top=2.54cm,bottom=2.54cm]{geometry}

\usepackage[boxed, linesnumbered,noend]{algorithm2e}
\usepackage{adjustbox}
\usepackage{eurosym}
\usepackage{lscape}
\usepackage{nicefrac}
\usepackage{array}
\usepackage{paralist}
\usepackage{authblk}
\usepackage{graphicx}
\usepackage{booktabs}
\usepackage{placeins}
\usepackage{amsmath}
\usepackage{mathtools}
\usepackage{bm}
\usepackage{amssymb}
\usepackage{color}
\usepackage[acronym]{glossaries}
\usepackage{pgfplots}
\usepackage[normal]{threeparttable}
\usepackage{ulem}
\usepackage{natbib}
\usepackage[titletoc,title]{appendix}
\usepackage[hyphens]{url} 
\usepackage{ellipsis}
\usepackage{breqn} 
\usepackage{chngcntr}  
\usepackage{enumerate}
\usepackage{theorem}
\usepackage{etoolbox}
\usepackage{multirow}
\usepackage{colortbl}
\usepackage[aboveskip=2pt,belowskip=2pt,format=plain,textformat=simple,textfont=footnotesize,labelfont={bf,footnotesize},indention=1cm]{caption}
\usepackage{subcaption}
\usepackage{rotating}
\usepackage{tikz}
\usetikzlibrary{arrows.meta}
\usetikzlibrary{backgrounds}
\usepackage{pgfplotstable}

\makeglossaries
\newtheorem{proofU}{Proof}
\newtheorem{observation}{Observation}
\newtheorem{example}{Example}
\newtheorem{definition}{Definition}
\newtheorem{theorem}{Theorem}
\newtheorem{corollary}{Corollary}
\newtheorem{result}{Result}

\begin{document}
% first put your acronyms and abbrevs.
%\input{content/abbreviations}
%\input{content/rwth-colors}
\input{./chapters/acronyms.tex}
%% then the title
\input{chapters/titlepage.tex}
%% and your chapters
\newpage

\input{./chapters/introduction.tex}
\input{./chapters/problem.tex}
\input{./chapters/mechanisms.tex}
\input{./chapters/properties.tex}
\input{./chapters/experiments.tex}
\input{./chapters/results.tex}
\input{./chapters/winwin.tex}
\input{./chapters/conclusion.tex}

\begin{appendices}
	\input{./chapters/appendix.tex}
\end{appendices}

%% finally the bib file
\singlespacing{
\bibliographystyle{model5-names}%\biboptions{authoryear}
\bibliography{./main,./oscm} } % if more than one, comma separated
\end{document}

%% file: chapters/acronyms.tex
% a
% b
% c
\newacronym{abk:ccp}{CCP}{customer-centered pooling}
% d
% e
% f
% g
% h
% i
% j
% k
% l
% m
\newacronym{abk:msp}{MSP}{mobility service provider}
% n
% o
% p
\newacronym{abk:pcp}{PCP}{provider-centered pooling}
% q
% r
% s
\newacronym{abk:sol}{SRO}{solitary rides only}
% t
% u
% v
% w
% x
% y
% z	
%%%%%%%%%%%%%%%%%%%%%%%%%%%%%%%%%%%%%%%%%%%%%%%%%%%%%%%%%%%%%%%%%%%%%%%%%%%%%%%%%%%%%%%%%%%%%%%%%%%%%%%%%%%%%%%%%%%%%%%%%%
% Commands
% \newcommand{}{}
%-Large-Arabic-------------------------------------------------------------------------------
% A
%\newcommand{}{A}
% B
%\newcommand{}{B}
% C
%\newcommand{}{C}
% D
%\newcommand{}{D}
\newcommand{\drop}[1]{DO(#1)}
% E
%\newcommand{}{E}
% F
%\newcommand{}{F}
% G
%\newcommand{}{G}
\newcommand{\graph}{G}
% H
%\newcommand{}{H}
% I
%\newcommand{}{I}
% J
%\newcommand{}{J}
% K
%\newcommand{}{K}
% L
%\newcommand{}{L}
% M
%\newcommand{}{M}
% N
%\newcommand{}{N}
% O
%\newcommand{}{O}
% P
%\newcommand{}{P}
\newcommand{\pu}[1]{PU(#1)}
% Q
%\newcommand{}{Q}
% R
%\newcommand{}{R}
\newcommand{\rec}[1]{REC(#1)}
% S
%\newcommand{}{S}
% T
%\newcommand{}{T}
% U
%\newcommand{}{U}
% V
%\newcommand{}{V}
% W
%\newcommand{}{W}
% X
%\newcommand{}{X}
% Y
%\newcommand{}{Y}
% Z
%\newcommand{}{Z}
%-Small-Arabic-------------------------------------------------------------------------------
% a
%\newcommand{}{a}
\newcommand{\arcIndex}{a}
% b
%\newcommand{}{b}
% c
%\newcommand{}{c}
\newcommand{\cost}[1]{c(#1)}
\newcommand{\costSol}[1]{c^{\text{S}}(#1)}
\newcommand{\costccp}[1]{c^{\text{CC}}(#1)}
\newcommand{\costpcp}[1]{c^{\text{PC}}(#1)}
% d
%\newcommand{}{d}
\newcommand{\destination}{d}
% e
%\newcommand{}{e}
% f
%\newcommand{}{f}
% g
%\newcommand{}{g}
% h
%\newcommand{}{h}
% i
%\newcommand{}{i}
% j
%\newcommand{}{j}
% k
%\newcommand{}{k}
% l
%\newcommand{}{l}
\newcommand{\locTaxi}[2]{\ell\left(#1,#2\right)}
% m
%\newcommand{}{m}
% n
%\newcommand{}{n}
% o
%\newcommand{}{o}
\newcommand{\origin}{o}
% p
%\newcommand{}{p}
\newcommand{\pay}[1]{p(#1)}
\newcommand{\paysol}[1]{p^{\text{S}}(#1)}
\newcommand{\paypcp}[1]{p^{\text{PC}}(#1)}
\newcommand{\payccp}[1]{p^{\text{CC}}(#1)}
\newcommand{\variablePrice}[2]{p^{\text{v}}(#1,#2)}
\newcommand{\fixedPrice}{p^{\text{f}}}
\newcommand{\changeFee}{p^{\text{c}}}
% q
%\newcommand{}{q}
% r
%\newcommand{}{r}
% s
%\newcommand{}{s}
% t
%\newcommand{}{t}
\newcommand{\pickupTime}{t^{\text{o}}}
\newcommand{\dropoffTime}{t^{\text{d}}}
% u
%\newcommand{}{u}
% v
%\newcommand{}{v}
\newcommand{\valueofTime}{v}
% w
%\newcommand{}{w}
\newcommand{\weightLength}{w^l}
\newcommand{\weightTime}{w^t}
% x
%\newcommand{}{x}
% y
%\newcommand{}{y}
% z
%\newcommand{}{z}
%-Large-Calligraphic-------------------------------------------------------------------------
% A
%\newcommand{}{\mathcal{A}}
\newcommand{\setofArcs}{\mathcal{A}}

% B
%\newcommand{}{\mathcal{B}}
% C
%\newcommand{}{\mathcal{C}}
\newcommand{\setofCustomers}{\mathcal{C}}
% D
%\newcommand{}{\mathcal{D}}
% E
%\newcommand{}{\mathcal{E}}
% F
%\newcommand{}{\mathcal{F}}
% G
%\newcommand{}{\mathcal{G}}
% H
%\newcommand{}{\mathcal{H}}
% I
%\newcommand{}{\mathcal{I}}
% J
%\newcommand{}{\mathcal{J}}
% K
%\newcommand{}{\mathcal{K}}
% L
%\newcommand{}{\mathcal{L}}
\newcommand{\setofNodes}{\mathcal{L}}
% M
%\newcommand{}{\mathcal{M}}
% N
%\newcommand{}{\mathcal{N}}
% O
%\newcommand{}{\mathcal{O}}
% P
%\newcommand{}{\mathcal{P}}
% Q
%\newcommand{}{\mathcal{Q}}
% R
%\newcommand{}{\mathcal{R}}
% S
%\newcommand{}{\mathcal{S}}
% T
%\newcommand{}{\mathcal{T}}
% U
%\newcommand{}{\mathcal{U}}
% V
%\newcommand{}{\mathcal{V}}
\newcommand{\setofVehicles}{\mathcal{U}}
% W
%\newcommand{}{\mathcal{W}}
% X
%\newcommand{}{\mathcal{X}}
% Y
%\newcommand{}{\mathcal{Y}}
% Z
%\newcommand{}{\mathcal{Z}}
%-Large-Greek--------------------------------------------------------------------------------
% Alpha
%\newcommand{}{A}
% Beta
%\newcommand{}{B}
% Gamma
%\newcommand{}{\Gamma}\\
% Delta
%\newcommand{}{\Delta}
\newcommand{\deviation}{\Delta}
% Epsilon
%\newcommand{}{E}
% Zeta
%\newcommand{}{Z}
% Eta
%\newcommand{}{H}
% Theta
%\newcommand{}{\Theta}
% Iota
%\newcommand{}{I}
% Kappa
%\newcommand{}{K}
% Lambda
%\newcommand{}{\Lambda}
% Mu
%\newcommand{}{M}
% Nu
%\newcommand{}{N}
% Xi
%\newcommand{}{\Xi}
% Omicron
%\newcommand{}{O}
% Pi
%\newcommand{}{\Pi}
% Rho
%\newcommand{}{P}
% Sigma
%\newcommand{}{\Sigma}
% Tau
%\newcommand{}{T}
% Upsilon
%\newcommand{}{\Upsilon}
%\newcommand{\vertTransFunc}[1]{\Upsilon\left(#1\right)}
% Phi
%\newcommand{}{\Phi}
\newcommand{\sigAmount}{\Phi}
% Chi
%\newcommand{}{X}
% Psi
%\newcommand{}{\Psi}
% Omega
%\newcommand{}{\Omega}
%-Small-Greek--------------------------------------------------------------------------------
% alpha
%\newcommand{}{\alpha}
% beta
%\newcommand{}{\beta}
% gamma
%\newcommand{}{\gamma}
\newcommand{\opCost}{\gamma}
% delta
%\newcommand{}{\delta}
\newcommand{\discount}{\delta}
% epsilon
%\newcommand{}{\varepsilon}
% zeta
%\newcommand{}{\zeta}
% eta
%\newcommand{}{\eta}
% theta
%\newcommand{}{\vartheta}
% iota
%\newcommand{}{\iota}
% kappa
%\newcommand{}{\kappa}
% lambda
%\newcommand{}{\lambda}
% mu
%\newcommand{}{\mu}
% nu
%\newcommand{}{\nu}
% xi
%\newcommand{}{\xi}
% omicron
%\newcommand{}{o}
% pi
%\newcommand{}{\pi}
% phi
\newcommand{\relSave}{\phi}
% rho
%\newcommand{}{\rho}
\newcommand{\poolBool}{\rho}
% sigma
%\newcommand{}{\sigma}
\newcommand{\sched}[1]{\sigma\left(#1\right)}
\newcommand{\schedPast}[2]{\overleftarrow{\sigma}^{#2}\left(#1\right)}
\newcommand{\schedFuture}[2]{\overrightarrow{\sigma}^{#2}\left(#1\right)}
% tau
%\newcommand{}{\tau}
\newcommand{\taxi}{\tau(\trip)}
% upsilon
%\newcommand{}{\upsilon}
% phi
%\newcommand{}{\phi}
\newcommand{\tripCustomers}{\phi(\trip)}
% chi
%\newcommand{}{\chi}
% psi
%\newcommand{}{\psi}
% omega
%\newcommand{}{\omega}
\newcommand{\maxWait}{\omega}
%-Other-------------------------------------------------------------------
\newcommand{\arc}{(i,j)}
\newcommand{\ifM}{\text{if  }}
\newcommand{\elseM}{\text{else}}
% more stuff
\newcommand{\summe}[1]{\sum_{\scriptstyle\mathclap{#1}}}
\newcommand{\blue}[1]{\textcolor{darkblue}{#1}}
\newcommand{\red}[1]{\textcolor{red}{#1}}

%% file: chapters/titlepage.tex
%----------------------------------------------------------------------------------
% titlepage template
% author: Maximilian Schiffer
% version: 1.0 / 11.05.2017
%----------------------------------------------------------------------------------
\title{\large The Customer is Always Right:\\ Customer-Centered Pooling for Ride-Hailing Systems}

% and the authors
\author[1]{\normalsize Paul Karaenke}
\author[2,3]{\normalsize Maximilian Schiffer}
\author[4]{\normalsize Stefan Waldherr}

\small
\affil[1]{\footnotesize Department of Informatics, Technical University of Munich, Garching D-85748, Germany}
\affil[2]{\footnotesize TUM School of Management, Technical University of Munich, Munich D-80333, Germany}
\affil[3]{\footnotesize Munich Data Science Institute, Technical University of Munich, Garching D-85748, Germany}
\affil[4]{\footnotesize School of Business and Economics, Vrije Universiteit Amsterdam, Amsterdam 1081HV, The Netherlands}

\affil[ ]{
	\scriptsize
	paul.karaenke@tum.de,
	schiffer@tum.de,
	s.m.g.waldherr@vu.nl}

\date{}

\maketitle
% finally your abstract
\begin{abstract}
\begin{singlespace}
{\small\noindent Today's ride-hailing systems experienced significant growth and ride-pooling promis\-es to allow for efficient and sustainable on-demand transportation. However, efficient ride-pooling requires a large pool of participating customers. To increase the customers' willingness for participation, we study a novel customer-centered pooling (CCP) mechanism, accounting for individual customers' pooling benefits. We study the benefit of this mechanism from a customer, fleet operator, and system perspective, and compare it to existing provider-centered pooling (PCP) mechanisms. We prove that it is individually rational and weakly dominant for a customer to participate in CCP, but not for PCP. We substantiate this analysis with complementary numerical studies, implementing a simulation environment based on real-world data that allows us to assess both mechanisms' benefit. To this end, we present results for both pooling mechanisms and show that pooling can benefit all stakeholders in on-demand transportation. Moreover, we analyze in which cases a CCP mechanism Pareto dominates a PCP mechanism and show that a mobility service operator would prefer CCP mechanisms over PCP mechanisms for all price segments. Simultaneously, CCP mechanisms reduce the overall distance driven in the system up to 32\% compared to not pooling customers.	Our results provide decision support for mobility service operators that want to implement and improve pooling mechanisms as they allow us to analyze the impact of a CCP and a PCP mechanism from a holistic perspective. Among others, we show that CCP mechanisms can lead to a win-win situation for operators and customers while simultaneously improving system performance and reducing emissions.

\smallskip}
\end{singlespace}
\end{abstract}

%% file: chapters/introduction.tex
\section{Introduction}\label{sec:introduction}
In recent years, ride-hailing services extended the sharing economy to taxi services and experienced severe growth, especially in metropolitan areas. In the US, the number of ride-hailing trips showed an annual growth rate of 150\% between 2013 and 2018 and more than two billion rides per year \citep{BalticCappyEtAl2019}. The ride-hailing sector continues its growth with a revenue growth rate of 4.13\% for the year 2019. While experts envisioned ride-hailing services to remedy harm caused by local externalities of traffic phenomena such as congestion and noxious emissions, recent dynamics revealed contrary effects. In some areas, congestion and emission levels became even worse, not least because of induced demand from ride-hailing systems: in Manhattan, the number of for-hire vehicles exploded from 47,000 vehicles in 2013 to 103,000 vehicles in 2018, 68,000 used for ride-hailing services.  During this period, a 13\% drop in average traffic speed from 6.5 mph to 4.7 mph occurred \citep{Hu2017}. Accordingly, today's ride-hailing systems face the same criticism as private transport; experts claim that, besides induced demand, low utilization is a major obstacle.

To this end, ride-pooling concepts may allow for higher vehicle utilization. Herein, a \gls{abk:msp} that offers a ride-hailing service tries to match passengers whose rides (partially) overlap and schedules them on the same vehicle. In exchange for resulting inconveniences, passengers receive a price discount. Ideally, passengers benefit from a discounted transport fare while \glspl{abk:msp} receive an increased revenue due to operational cost savings. Implemented in the right way and accepted by a sufficient number of customers, such ride-pooling services bear the potential for a win-win situation that increases utilization in today's ride-hailing systems.

Uber and Lyft, two of the largest ride-hailing providers worldwide,  launched ride-pooling options such as Uber Pool or Lyft Line, which yield the above mentioned win-win situation, decreasing cost at the passenger side and increasing revenues at the mobility side, in theory. In practice, several obstacles remain. \glspl{abk:msp} often lose money as passengers receive a discount for their willingness to be matched on a shared ride, independent of a successful matching \citep{CBI2020}. Customers often perceive a maximum loss of comfort when selecting the pooled ride due to inadequate but yet economically worthwhile matchings. These effects relate to poor matchings of customer trips, mostly due to two reasons: first, the share of customers willing to accept a pooled ride is too low to allow for constantly good matchings. Second, existing matching mechanisms operate mostly on an economically objective, considering few side constraints to limit the customers' additional discomfort \citep[cf.][]{Alonso-MoraSamaranayakeEtAl2017}. In such a chicken and egg dilemma, customer willingness to accept a pooling offer may decrease due to a bad customer experience, further worsening the probability of finding a suitable matching in a reduced customer pool.

Against this background, we study the impact of customer-centered ride-pooling. To set our study apart from recent work, we detail the status quo and current challenges~(Section~\ref{subsec:statusQuo}), before we review related work~(Section~\ref{subsec:stateoftheart}), and state our contribution (Section~\ref{subsec:scope}), as well as the paper's organization~(Section~\ref{subsec:structure}).

%---------------------------------------------------------------
\subsection{Status Quo \& Challenges}\label{subsec:statusQuo}
Currently, major mobility providers offer a \textit{fixed-discount tariff}, i.e., provide an up-front discount to passengers in exchange for an option to include them into a pooled ride. Each passenger who opted to accept pooled rides receives a discount independently of being matched with another customer in practice. In this case, the provider takes the risk of a reduced profit margin in case of unmatched rides and chooses a \textit{\gls{abk:pcp}} mechanism. Here, the provider pools customers with the sole objective to maximize its profit under certain constraints. Typical constraints that are nowadays applied in practice and studied in the literature are maximum delays or maximum detours \citep{HosniEtAl2014,KeEtAl2020}.

The efficiency of matching passenger requests to pooled rides depends significantly on the number of customer requests available in the matching. Hence, successfully sustaining pooling systems requires win-win situations in which both the \gls{abk:msp} and customers profit, such that more potential customers opt for pooled rides. Current practice fosters win-lose situations either between the \gls{abk:msp} and the customers or in between customers: offering a fixed-discount tariff, the \gls{abk:msp} takes the risk of a reduced profit due to unmatched requests. Accordingly, it stretches constraints on delay times or detours to a maximum to reduce its risk. Although perceived as still acceptable from the \gls{abk:msp} perspective, these boundaries may often go beyond a customer's bottom line for acceptable loss in comfort. Often, the customer wins a cheap solitary ride, and the \gls{abk:msp} loses profit, or the \gls{abk:msp} matches customers whose losses in comfort and time outweigh their monetary savings. Additional win-lose situations likely arise between pooled customers: the fixed discount often does not reflect the comfort loss of both customers as their relative detours or waiting times may differ. In total, such win-lose situations may entail a drop instead of an increase in the customers' willingness to volunteer in ride-pooling as negative experiences accumulate over time. Indeed, several reports indicate that customers are not satisfied with the ``lottery'' nature of the resulting mechanism \citep[see, e.g.,][]{MorrisEtAl2020,PrattEtAl2019}.

A \textit{flexible-discount tariff} that accounts for each customer's individual preference and her contribution to the pooled ride may improve the customer experience and the individual acceptance of comfort loss. Such a \textit{\gls{abk:ccp}} considers each customer's value of time in the objective and establishes a transparent and fair matching that may increase a customer's willingness to participate in pooling. In practice, such a mechanism can be implemented by asking for each customer's preference when making a trip request. For example, a drop-down menu can be added to existing ride-pooling apps in which customers select their value of time directly or select from a discrete set of ``urgency'' categories, which are then internally converted into respective monetary values. Studying such a mechanism is the scope of this paper.

%---------------------------------------------------------------
\subsection{State of the Art}\label{subsec:stateoftheart}
Our work relates to two different literature streams, ride-sharing, and ride-pooling, which we concisely review in the following. We note that the terminology of ride-sharing and ride-pooling, which we explicitly separate, is often used synonymously in existing works.

\paragraph{Ride-Sharing:}
In classical ride-sharing programs, two (or more) people who plan to travel similar routes can share a car and split the resulting cost. Here, one participant acts as a driver and picks up and drops off other participants (riders) along her way. A central platform matches participants for a fee but does not provide a dedicated driver to satisfy customer requests. Early versions of ride-sharing required participants to announce their requests well in advance \citep[cf.][]{BerbigliaCordeauEtAl2010}, but advances in mobile internet technologies paved the way for dynamic ride-sharing programs \citep[cf.][]{FuruhataDessoukyEtAl2013}. 

Many studies focus on the technical aspects of the underlying matching mechanisms in this field, aiming to find matchings and cost-sharing schemes for customers, considering their preferences in a two-sided matching market. \citet{WangAgatzEtAl2017} focused on stable two-sided matchings in dynamic ride-sharing systems, while \citet{PengShanEtAl2020} focused on stable matchings, including payment design options. \citet{RasulkhaniChow2019} study a route cost assignment game to analyze the market equilibrium. Some papers focused on fair cost allocations for ride-sharing matchings \citep{WangYangEtAl2018,LuQuadrifoglio2019,PengShanEtAl2020}, but do not account for individual customer preferences. \citet{QianZhangEtAl2017} focused on pricing aspects combined with a matching problem, analyzing incentives for optimally assigning ride-sharing matches. Moreover, some auction-based mechanisms to coordinate ride-sharing systems exist \citep[cf.][]{BianLiu2019}. These works mainly apply the Vickrey-Clarke-Groves mechanism to a ride-pooling setting to maximize social welfare. 

So far, none of the existing works accounts for individual customer preferences, and the general problem setting of a ride-sharing system is only loosely related to our study. We refer to \citet{AgatzEreraEtAl2012,FuruhataDessoukyEtAl2013}, and \citet{HoSzetoEtAl2018} for comprehensive overviews of such (dynamic) ride-sharing systems. 

\paragraph{Ride-Pooling:}
In ride-hailing, an \gls{abk:msp} provides a dedicated driver to serve customers' requests. The \gls{abk:msp}s can either operate a centrally coordinated fleet of vehicles and drivers (e.g., a taxi service) or rely on decentralized ride-sourcing (e.g., Uber, Lyft). \citet{WangYang2019} and \citet{FengEtAl2020} provide state-of-the-art summaries of current ride-hailing systems, the former focusing in particular on ride-sourcing. Ride-pooling extends ride-hailing to the concept o ride-sharing and allows shared vehicles to transport customers with similar routes. In contrast to classical ride-sharing, passengers are picked up and dropped off by the dedicated driver provided by the \gls{abk:msp}. We refer the reader to \citet{VazifehSantiEtAl2018} for a general impact analysis of ride-hailing fleets, to \citet{QiLiEtAl2018} for an overview on shared mobility in last-mile deliveries, and to \citet{HoSzetoEtAl2018} for a recent overview on classical and emerging dial-a-ride problem studies.

In this field, research on ride-pooling focused mostly on optimizing vehicles' utilization, either by reducing the fleet size or lowering operational cost \citep{HosniEtAl2014,Alonso-MoraSamaranayakeEtAl2017}. Within this domain, pricing strategies mostly capture ride-sourcing applications, simultaneously optimizing prices offered for non-pooled and pooled rides that maximize the \gls{abk:msp}s revenue \citep{YanZhuEtAl2020}, or focused on spatial pricing for drivers  \citep{BimpikisCandoganEtAl2019}.

Several studies addressed ride-sharing systems' operational aspects, often from classical transportation, vehicle routing, or matching perspectives. \citet{ChenMesEtAl2019} analyzed a ride-sharing system with meeting points and return restrictions, while \citet{AgussurjaChengEtAl2019} focused on a stochastic ride-sharing model based on a two-stage Markov decision process. \citet{YangQinEtAl2020} analyzed the impact of matching constraints such as a detour radius, and \citet{YangRenEtAl2020} focused on saturation effects between supply and demand in a ride-sharing system.  \citet{ChenWang2018} focused on maximizing ride-pooling systems' social welfare, optimizing vehicle capacities, and fleet size. \citet{LongTanEtAl2018} studied the impact of uncertain travel times on matchings in ride-pooling systems. 

To the best of our knowledge, no study exists that focuses on general customer-centered online-matching mechanisms that account for individual customer cost functions and incorporate each customer's inconvenience to allow for fair matchings and cost allocations in a ride-pooling system. 
%---------------------------------------------------------------
\subsection{Contribution}\label{subsec:scope}
We introduce the first study that evaluates the economic potential of \gls{abk:ccp} mechanisms for ride-hailing fleets based on a formal analysis of the discussed mechanisms and a detailed numerical study to evaluate their operational impact. We propose a \gls{abk:ccp} mechanism that can be implemented in practice by asking for a each customer's preference when making a trip request and allows for a fair matching in ride-pooling systems: customers receive a fair cost and discount allocation, reflecting each customer's individual preferences based on her value of time; operators must only grant a  discount in case of a successful pooling of customers. Specifically, our contributions are as follows. First, we prove that participating in a \gls{abk:ccp} mechanism is a weakly dominant strategy for customers. Contrary, \gls{abk:pcp} mechanisms may lead to unprofitable matchings for customers or missed matchings that could have been profitable for both customers and \glspl{abk:msp}. We then implement the proposed pooling mechanisms in a simulation environment to conduct an extensive numerical study based on a real-world data set. We analyze each mechanism's impact on the customer's cost, the operator's profit, fleet utilization, and its environmental impact. We synthesize several managerial insights by combining findings from our formal analysis and our numerical results. Finally, we provide evidence that \gls{abk:ccp} mechanisms are preferable over \gls{abk:pcp} mechanisms for both operators and customers while leading to significant reductions in carbon dioxide emissions.
%---------------------------------------------------------------
\subsection{Structure}\label{subsec:structure}
The remainder of this paper is as follows. Section~\ref{sec:problem_setting} introduces our problem setting, while Section~\ref{sec:methodology} presents our methodology, including a formal analysis of each matching mechanism. We detail our numerical study and the design of experiments in Section~\ref{sec:experiments} and discuss the impact of the studied pooling mechanisms in Section~\ref{sec:results}. We synthesize these results in Section \ref{sec:synthesis} to allow for a high-level assessment. 
Section~\ref{sec:conclusion} concludes this paper with a short summary.

%% file: chapters/problem.tex
\section{Problem Setting}\label{sec:problem_setting}
We consider mobility market places where \glspl{abk:msp} interact with customers via an online platform, e.g., a smartphone app, and offer ride-hailing services to customers. Using this platform, the \glspl{abk:msp} extract necessary customer information and set prices for solitary and pooled rides. Customers request the \glspl{abk:msp}' services and choose a solitary ride or volunteer to be pooled. In case a customer volunteers for pooling, she indicates her value of time, e.g., by selecting a respective category in the app when submitting her request. If a customer volunteers to be pooled (i.e., she is \textit{poolable}), an \gls{abk:msp} can match her request with another customer's request to a pooled ride, served by a single vehicle to reduce costs and to increase the fleet's efficiency. In return, the \gls{abk:msp} compensates poolable customers with reduced prices compared to solitary rides. In this context, we focus on the effect of matching mechanisms that \glspl{abk:msp} may use to form pooled rides. Accordingly, we abstract from the multiple \gls{abk:msp} setting in the following discussion and refer to a single \gls{abk:msp}, i.e., its matching mechanism. Clearly, this does not limit our results to a single \gls{abk:msp} setting as all \glspl{abk:msp} in such a system may use the analyzed mechanisms.
%--
\subsection{Assumptions} First, we assume that all participants behave economically rational, aiming to minimize their costs; the \gls{abk:msp} always assigns cost-optimal routes, and customers prefer their cost-minimal option. This is in line with recent studies in the field of mobility as a service. Accordingly, we impose a customer's maximum waiting time as a hard constraint. However, customer cost functions are not subject to any conditions except that the value of time is non-negative. Therefore, our framework allows to include maximum waiting times as a soft constraint, incorporated in a customer's cost function.
Second, we assume that all customer requests arrive online and require immediate pick-up. Third, we allow only two customers to share a ride at once. Both assumptions reflect current best practices that keep passenger discomfort limited. Fourth, we neglect fixed costs of the \gls{abk:msp} as they do not affect short-term operational decisions.
%--
\subsection{Model}
The realization of rides takes place on a road network, represented as a graph $\graph = (\setofNodes,\setofArcs)$ with a set of nodes $\setofNodes$ and a set of arcs $\setofArcs$. Each node $l\in\setofNodes$ represents an origin or destination location of a customer trip; each arc  $(l,l')\in\setofArcs$ denotes a shortest path between nodes $l,l'\in\setofNodes$. 

For each customer $i \in \setofCustomers$, we define a customer trip \textit{request} as a tuple $r_i = (\origin_i,\destination_i,\pickupTime_i,\valueofTime_i,\maxWait_i,\poolBool_i)$, consisting of the customer's origin $\origin_i\in\setofNodes$, her destination $\destination_i\in\setofNodes$, her preferred pick-up time $\pickupTime_i$, her individual value of time $\valueofTime_i\in \mathbb{R}_{\ge 0}$, a maximum waiting time $\maxWait_i$, and a binary parameter $\poolBool_i\in\{0,1\}$ indicating whether a customer participates in pooling ($\poolBool_i = 1$) or not ($\poolBool_i= 0$). If a customer's maximum waiting time $\maxWait_i$ is exceeded, she cancels the travel request, i.e., a customer only accepts a ride offer with a waiting time below $\maxWait_i$ if all other constraints are met. As we assume immediate pickup requests, the preferred pickup time of a customer equals the time the request is made.

Customer $i$'s costs comprise a payment $\pay{i}$ to the \gls{abk:msp} in exchange for the offered service and the cost for the customer's time spent. This cost depends on the time span between a customer's preferred pick up time at her origin ($\pickupTime_i$) and the time when she is dropped at her destination~($\dropoffTime_i$), as well as on her individual value of time $\valueofTime_i$. Hence, both waiting and traveling times induce costs for a customer $i$ based on $\valueofTime_i$, defining her valuation for a time unit spent; i.e., these costs of time account for the time between requesting a ride and arriving at her destination. Then, a customer's cost function reads 
\begin{equation}
\cost{i} =  \pay{i} + \valueofTime_i \cdot \left( \dropoffTime_i  - \pickupTime_i \right).
\end{equation}

The \gls{abk:msp} operates a fleet of vehicles $u\in\setofVehicles$ that serve customers' transportation requests. Here, the \gls{abk:msp} receives a fare for each served ride and aims to maximize its profit while accounting for milage dependent cost.
The \gls{abk:msp} matches customers to vehicles over a time horizon $\mathcal{T}$. Whenever a new customer request $r_i$ appears at time $t\in\mathcal{T}$, the \gls{abk:msp} assigns the customer to either an empty vehicle or to a (potentially) ongoing ride in case that both customers volunteer for pooling. 

To describe the operational state of the \gls{abk:msp}'s fleet over $\mathcal{T}$, we use a schedule $\sched{u}$ for each vehicle, which is an ordered list of quadruples. Each quadruple $(l_k,t_k,o_k,c_k)$ links a vehicle's operation $o_k$ to a customer $c_k$, location $l_k$, and time $t_k$. We order a schedule such that the time component of each succeeding quadruple is increasing. A vehicle can perform three operations ($o_k \in \{PU,DO,REC\}$): when a vehicle picks a customer up ($o_k = PU$), $l_k\in\setofNodes$ denotes a customer's pickup location and $t_k$ denotes the currently scheduled (expected) pickup time. When a vehicle drops a customer off ($o_k = DO$), $l_k\in\setofNodes$ denotes a customer's drop-off location and $t_k$ denotes the currently scheduled (expected) drop-off time. A vehicle receives a new request when the \gls{abk:msp} assigns a request to the vehicle ($o_k = REC$). Here, $l_k$ and $t_k$ state the time and vehicle location at the point of assignment such that $l_k$ can be either a location ($l_k\in\setofNodes$) or a road segment ($l_k\in\setofArcs$).

Assigning a request to a vehicle requires scheduling all three operations at once: the vehicle's schedule needs to be recalculated such that it contains all operations for a customer in the order $REC\rightarrow PU\rightarrow DO$. We use $\schedPast{u}{t}$ to refer to the inactive part of vehicle $u$'s schedule at time $t\in\mathcal{T}$, i.e., the operations scheduled at $t_k < t$ that vehicle $u$ processed up to time $t$ . Analogously, $\schedFuture{u}{t}$ denotes the active part of schedule $\sched{u}$, i.e., scheduled operations with $t_k \ge t$ that have been assigned to vehicle $u$ but not been processed up to time $t$.

Whenever we assign a customer to a vehicle at time $t$, we update its schedule $\sched{u}$ and include the respective operations in $\schedFuture{u}{t}$. This update may require updating already existing entries in $\schedFuture{u}{t}$, e.g., if picking up a new customer is included in between the pick-up and the drop-off of an already active customer. Example~\ref{exp:schedule} details the connection between the different activities and the mechanism of assigning a request to a vehicle.

\begin{example}\label{exp:schedule}
	We consider a customer $i$, her request $r_i$,  and two vehicles $u_1,u_2$. Let $u_1$ be a currently empty vehicle with no active schedule that is at location $l_1$. If customer $i$ is assigned to $u_1$, its active schedule $\schedFuture{u_1}{\pickupTime_i}$ is changed to $\left[(l_1,\pickupTime_i,REC,i),(\origin_i,\tilde{t}_{1,1},PU,i),(\destination_i,\tilde{t}_{1,2},DO,i)\right]$ where $\tilde{t}_{1,1}$ and $\tilde{t}_{1,2}$ result from the shortest paths from  $l_1$ to $\origin_i$ and $\origin_i$ to $\destination_i$, respectively.
	
	Let $u_2$ be a vehicle which is currently at location $l_2$ and has an active schedule of $\schedFuture{u_2}{\pickupTime_i} = \left[(\origin_j,t_{2,1},PU,j),(\destination_j,t_{2,2},DO,j)\right]$ where $j$ is another customer that has requested a ride prior to~$i$. Customer $i$ might be assigned to $u_2$ by altering the future schedule of $u_2$ in the following way:
	 $\left[(l_2,\pickupTime_i,REC,i),(\origin_j,\tilde{t}_{2,1},PU,j),(\origin_i,\tilde{t}_{2,2},PU,i),(\destination_i,\tilde{t}_{2,3},DO,i),(\destination_j,\tilde{t}_{2,4},DO,j)\right]$, where $\tilde{t}_{2,1},\ldots,\tilde{t}_{2,4}$ are again determined by calculating the shortest paths between the respective locations of the activities. In this case, customer $j$ is picked up before customer $i$ and dropped off after $i$ is dropped off. Hence, $i$ and $j$ share a vehicle between $\origin_i$ and $\destination_i$.
\end{example}

%% file: chapters/mechanisms.tex
\section{Methodology}\label{sec:methodology}
In the following, we first present different assignment mechanisms before we formally discuss their properties and detail the flexibility of our \gls{abk:ccp} mechanism.
%--
\subsection{Mechanisms}
We consider three different mechanisms, a \gls{abk:pcp} and a \gls{abk:ccp} mechanism, and a solitary rides only option that serves as a baseline. These mechanisms are described in the following.
\paragraph{Solitary Rides Only:}
An \gls{abk:msp} may offer only solitary rides, matching customers solely to empty vehicles. Then, the \gls{abk:msp} charges a payment $\paysol{i} = \fixedPrice + \variablePrice{\origin_i}{\destination_i}$ from customer $i$. Here, $\fixedPrice$ constitutes a basic charge, independent of a customer's origin and destination, while $\variablePrice{\origin_i}{\destination_i}$ constitutes a variable price component as a function of the customers origin-destination pair, which is based on the length of the shortest path between $\origin_i$ and $\destination_i$ and the time required to traverse the path.

\begin{definition}[Solitary Rides Only]
	In a \textit{\gls{abk:sol}} mechanism, the \gls{abk:msp} assigns vehicles in such a way that the total driving distance (including the distance between a vehicles location and the origin of the customer) is minimized, and a vehicle picks up each customer $i$ within her maximal waiting interval $\maxWait_i$ if possible. Customers are never pooled.
\end{definition}

%--
\paragraph{Provider-Centered Pooling:}
Offering discounts to customers in exchange for the customers' willingness to be poolable is the status quo in most ride-hailing fleets (cf. \cite{Wang2019}). Here, an \gls{abk:msp} pools customers among each other to increase its profit. However, a customer may not be pooled if this benefits the \gls{abk:msp}'s profit maximization.
The \gls{abk:msp} charges a payment $\paypcp{i} = \discount \cdot \paysol{i} = \discount \cdot \left(\fixedPrice + \variablePrice{\origin_i}{\destination_i}\right)$ from customer $i$, with $\discount \in (0,1]$ being the discount factor offered to the customer for her willingness to be pooled. 
Often, the \gls{abk:msp} imposes constraints on possible matchings to limit customer inconvenience. We consider such constraints by limiting the \textit{additional} duration of customer trips to a \textit{detour factor} $\deviation \in \mathbb{R}_{>0}$ that relates to the expected duration of solitary rides. 

\begin{definition}[Provider-Centered Pooling]
	In a \textit{\gls{abk:pcp}} mechanism, each poolable customer~$i$ pays $\discount \cdot \paysol{i}$. In exchange, the \gls{abk:msp} may pool $i$ with other customers as long as $i$'s trip is not prolonged by more than a factor $1+\deviation$ compared to $i$'s solitary ride.
\end{definition}

One of the main shortcomings of \gls{abk:pcp} is that the customer receives a discount independent of the outcome of the \gls{abk:msp}'s matching. Accordingly, one may see a customer's willingness to be pooled as a bet on receiving a free discount if no complementary ride exists. The \gls{abk:msp}'s discount offer remains a bet on receiving a sufficient complementary population of poolable customers, assuming that it can compensate all offered discounts by increased operational efficiency, which finally leads to an increase in profit. The \gls{abk:msp} may match any poolable customers dynamically while considering the $\deviation$ constraints, such that its costs are minimized and hence its profit is maximized. Therefore, in a \gls{abk:pcp} mechanism, only upper bounds on estimated times of arrival and customers' costs are known ex ante.
%--
\paragraph{Customer-Centered Pooling:}
As outlined above, only allowing for solitary rides or employing \gls{abk:pcp} may yield disadvantages and undesirable effects. To this end, we propose an alternative mechanism based on each customer's cost function. Here, the \gls{abk:msp} does no longer offer an unconditioned discount to poolable customers. Instead, the \gls{abk:msp} sets individual fares for pooled rides, which split between matched customers such that each customer receives a discount, e.g., proportional to her contribution to the necessary detours. 

Let $i$ be a poolable customer currently assigned to vehicle $u$ and let $j$ be a poolable customer that requests a ride at time $\pickupTime_j > \pickupTime_i$. Let $t^p_i$ and $t^p_j$ be the (expected) pick-up time of customers $i$ and $j$. To calculate the fare of a pooled ride in this setting, we distinguish between scenarios in which $i$ has or has not yet been picked up at the time that $u$ receives the request of $j$, i.e., if $t^p_i \leq \pickupTime_j$ or $\pickupTime_j < t^p_i$. Let $l$ be the location at which $u$ receives the request of $j$, i.e., the location of $u$ when its active schedule is changed to pool customers $i$ and $j$. Let $\dropoffTime_i,\dropoffTime_j$ be the respective expected drop-off times of customers $i$ and $j$. Then, the fare for a pooled ride results to 
\begin{multline}\footnotesize
\payccp{i,j} = \begin{cases}
\fixedPrice + \variablePrice{\origin_i}{l} +\variablePrice{l}{\origin_j} + \variablePrice{\origin_j}{\destination_i} + \variablePrice{\destination_i}{\destination_j} + \changeFee & \quad\text{if}\quad t_i^p \le \pickupTime_j \le t_j^p \leq \dropoffTime_i \leq \dropoffTime_j\\
\fixedPrice + \variablePrice{\origin_i}{l} +\variablePrice{l}{\origin_j} + \variablePrice{\origin_j}{\destination_j} + \variablePrice{\destination_j}{\destination_i} + \changeFee & \quad\text{if}\quad t_i^p \le \pickupTime_j \le t_j^p \leq \dropoffTime_j \leq \dropoffTime_i\\
\fixedPrice + \variablePrice{\origin_i}{\origin_j} + \variablePrice{\origin_j}{\destination_i} + \variablePrice{\destination_i}{\destination_j} + \changeFee & \quad\text{if}\quad \pickupTime_j < t_i^p \leq t_j^p \leq \dropoffTime_i \leq \dropoffTime_j\\
\fixedPrice + \variablePrice{\origin_i}{\origin_j} + \variablePrice{\origin_j}{\destination_j} + \variablePrice{\destination_j}{\destination_i} + \changeFee & \quad\text{if}\quad \pickupTime_j < t_i^p \leq t_j^p \leq \dropoffTime_j \leq \dropoffTime_i\\
\fixedPrice + \variablePrice{\origin_j}{\origin_i} + \variablePrice{\origin_i}{\destination_i} + \variablePrice{\destination_i}{\destination_j} + \changeFee & \quad\text{if}\quad \pickupTime_j < t_j^p \leq t_i^p \leq \dropoffTime_i \leq \dropoffTime_j\\
\fixedPrice + \variablePrice{\origin_j}{\origin_i} + \variablePrice{\origin_i}{\destination_j} + \variablePrice{\destination_j}{\destination_i} + \changeFee & \quad\text{if}\quad \pickupTime_j < t_j^p \leq t_i^p \leq \dropoffTime_j \leq \dropoffTime_i,
\end{cases}
\end{multline}
where $\changeFee$ is a fixed change fee for altering the vehicle schedule after the initial assignment of $i$.

We now recall that customer $i$ would pay $\paysol{i} = \fixedPrice + \variablePrice{\origin_i}{\destination_i}$ if she decided to take a solitary ride. In this case, her total cost result to $\costSol{i} = \paysol{i} + \valueofTime_i \cdot (t^{\text{d,S}}_i-\pickupTime_i)$, with $t^{\text{d,S}}_i$ being her drop-off time in the solitary case. Then, from the perspective of the total cost incurred, a \gls{abk:ccp} is beneficial for the coalition of customers $i$ and $j$ if and only if
\begin{multline}\footnotesize
\label{eq:profPool}
\hfill\costSol{i} + \costSol{j}  > \payccp{i,j} + \valueofTime_i  (t^{\text{d,CC}}_i -\pickupTime_i) + \valueofTime_j  (t^{\text{d,CC}}_j -\pickupTime_j),\hfill
\end{multline}
with $t^{\text{d,CC}}_i$ and $t^{\text{d,CC}}_j$ being the drop-off times for customers $i$ and $j$ for the pooled ride. If Equation~\ref{eq:profPool} holds true, there exists a division of $\payccp{i,j}$ among both customers such that $\payccp{i,j} = \payccp{i} + \payccp{j}$ and their individual costs for the pooled ride are lower than for taking a solitary ride. In this case, the \gls{abk:msp} implements the schedule for which such a division exists and determines payments such that both participants profit from being pooled.

\begin{definition}[Customer-Centered Pooling]
	In a \textit{\gls{abk:ccp}} mechanism, the \gls{abk:msp} may pool customers as long as it defines payments in such a way that each pooled customer $i$ incurs lower cost compared to taking a solitary ride, i.e., $\payccp{i} + \valueofTime_i  (t^{\text{d,CC}}_i -\pickupTime_i) \leq \costSol{i}\quad \forall i \in \setofCustomers$.
\end{definition}

The \gls{abk:ccp} mechanism allows for various pricing schemes to split payments between customers.  Indeed, any pricing scheme can be implemented as long as each customer's resulting total costs are at most equal to taking a solitary ride. In this case, the pooling decision is independent of the eventual division of payments (i.e., we obtain the same pooling, regardless of the payments). 
\begin{observation}
	In the \gls{abk:ccp} mechanism, cost sharing is independent from pooling decisions.
\end{observation}

We note that a customer can be pooled with two or more customers during the same trip if this is profitable. However, we do not allow more than two customers to be on the same vehicle at once. In addition, the probability of pooled rides decreases during the ride for customers, i.e., the closer customers get to their destination. For example, if a customer is one block away from her destination, a detour is unlikely to occur, since the possible savings for the shared portion of the remaining trip are unlikely to exceed the costs of the additional time required to pick up another customer. That is, if customer $i$ is in the taxi already, $t^{\text{d,CC}}_i$ and therewith $\valueofTime_i  (t^{\text{d,CC}}_i -\pickupTime_i)$ would increase to pick up an additional customer $j$, while $\costSol{i} + \costSol{j} - \payccp{i,j}$ decreases the closer $i$ is to her destination.

%Example~\ref{exp:ccpooling} details our 
%\begin{example}\label{exp:ccpooling}
%Consider the following ride pooling scenario from Manhattan (cf. Figure \ref{fig:ccpexample}) that demonstrates the CCP mechanism. At 3:00pm Alice requests a ride from Wall Street Street ($o_A$) in Lower Manhattan to 116th Street in East Harlem ($d_A$) with an expected arrival time of 3:35pm and a fare of \$24.50. At 3:10pm, Bob request a ride from 3rd Ave./34th St. ($o_B$) to Park Ave./83rd St. ($d_B$). A solitary ride would cost him \$15.00 and he would arrive at 3:36pm. However, the vehicle serving Alice can make a small detour to serve Bob's trip before bringing Alice to her destination. Then, Bob arrives at 3:42pm, while Alice's expected arrival time is delayed to 3:52pm. For the combined trip, the MSP charges \$30.50 which is shared between Alice and Bob. Assume that both, Alice and Bob, value each minute of their time with \$0.20, hence their cost for the additional travel time are \$3.40 for Alice and \$1.20 for Bob. However, instead of a sum of fares of \$39.50 they only have to pay a combined fare of \$30.50, saving \$9.00. In this case, decreasing Alice's fare to \$19.00 and Bob's fare to \$10.50 leaves both better off and decreases their total cost by \$2.20 each. Hence, it is profitable for both to share the vehicle and the proposed route is taken. 
%\end{example}

%% file: chapters/properties.tex
\subsection{Properties}\label{subsec:properties}
We now discuss properties of the pooling mechanisms and refer to Appendix~\ref{app:proofs} for all proofs. As a direct consequence of the definition of \gls{abk:ccp}, we find that if an \gls{abk:msp} offers \gls{abk:ccp} pooling, it is individually rational for a customer to volunteer for pooling. Setting oneself poolable cannot lead to a cost increase in case of a successful pooling.
\begin{theorem}\label{the:individualrational}
	In a \gls{abk:ccp} mechanism, it is individually rational for a customer to participate in pooling.
\end{theorem}
Indeed, for each poolable customer, neither actions of other customers (i.e., whether they set themselves poolable or not) nor the actions of the \gls{abk:msp} (i.e., which poolings and pricing mechanism it chooses) can lead to an outcome which is worse than being non-poolable. This allows for the following strengthened result.
\begin{corollary}\label{cor:weaklyDominant}
	In a \gls{abk:ccp} mechanism, it is a weakly dominant strategy for each customer to set herself poolable, regardless of other customers' or the \gls{abk:msp}'s actions.
\end{corollary}
Moreover, we observe that a \gls{abk:ccp} mechanism mitigates the \gls{abk:msp}'s risk to reduce fares without realizing a beneficial, cost-compensating matching.
\begin{observation}
	In a \gls{abk:ccp} mechanism, the \gls{abk:msp}  does no longer bear the risk for sunk costs that result from fare discounts granted for unsuccessful matchings.
\end{observation}

For customers who participate in \gls{abk:pcp}, volunteering for pooling is not always beneficial, because a customer's cost may be higher due to her detour, i.e., $\paysol{i} + \valueofTime_i  (t^{\text{d,S}}_i -\pickupTime_i) < \paypcp{i} + \valueofTime_i  (t^{\text{d,PC}}_i -\pickupTime_i) = \discount \cdot \paysol{i} + \valueofTime_i  (t^{\text{d,PC}}_i -\pickupTime_i)$ with $t^{\text{d,S}}_i$ and $t^{\text{d,PC}}_i$ being the drop-off times for \gls{abk:sol} and \gls{abk:pcp} respectively.
\begin{theorem}\label{the:notweakDom}
	Given a \gls{abk:pcp} mechanism, setting oneself poolable is not a weakly dominant strategy for any customer.
\end{theorem}
Indeed, Theorem~\ref{the:uprofitable} shows that regardless of the pricing structure of the \gls{abk:pcp} mechanism, its inflexibility may lead to welfare losses either for customers or for all participants.
\begin{theorem}\label{the:uprofitable}
	Fixed discounts and arrival time guarantees in \gls{abk:pcp} mechanisms can either lead to unprofitable poolings for customers or missed pooling opportunities that would have been profitable for both customers and the \gls{abk:msp}.
\end{theorem}

Our formal analysis suggests that from a customer's perspective, \gls{abk:ccp} is superior to \gls{abk:pcp} as it avoids unfair discount sharing. Indeed, various reports give evidence that in current \gls{abk:pcp} systems, customers gamble by setting themselves poolable, hoping to receive a discounted solitary ride and complain about inconvenience in case they lose the gamble and get pooled \citep[c.f.][]{MorrisEtAl2020}. Accordingly, one may expect more customers to participate in a \gls{abk:ccp} system compared to a \gls{abk:pcp} system in the long run. 

A \gls{abk:ccp} mechanism also offers advantages to the \gls{abk:msp} as discounts only apply to customers who actively contribute to the pooling. However, it remains an open question whether \gls{abk:ccp} or \gls{abk:pcp} allows for overall more efficient and profitable operations. In the remainder of this paper, we investigate this question through a profound numerical study.

\subsection{Pricing Flexibility of the Customer-Centered Pooling Mechanism}
\label{sec:flexPricing}
We recall that one advantage of the \gls{abk:ccp} mechanism is its flexibility in cost-sharing. As two customers are only pooled if the shared ride's total cost is lower than the sum of the solitary ride costs, there always exists a fare split such that both customers profit. Many cost-sharing mechanisms have been studied in the literature, e.g., dividing the profit equitable by using the Shapley value \citep{Shapley1953,WangAgatzEtAl2017}. An alternative approach to increase customers' willingness to participate in pooling is to maximize their chances for substantial savings. In the following, we present an integer program that allows for ex-post optimization to identify the maximum share of customers that profit from savings of at minimum $\sigAmount $. An \gls{abk:msp} may use such an ex-post optimum to implement a corresponding pricing scheme to meet customer preferences.

We define a \textit{run} $R$ of a vehicle as a part of its schedule that starts from the time when it picks up a single customer until being empty again. Let $\mathcal{R}$ be the set of runs. For each $R \in \mathcal{R}$, $\setofCustomers_R$ indicates the customers that are served during run $R$ and $p^R$ denotes the total fare charged by the \gls{abk:msp}. For each customer $i \in \setofCustomers$, $c^S_i$ denotes $i$'s total cost if she would take a solitary ride and $a_i$ denotes the cost of time that $i$ incurs as a result of being pooled. Let $p_i \in \mathbb{R}$ be the fare that $i$ has to pay, while $\relSave_i$ defines the relative cost savings for customer $i$ with price $p_i$, i.e., $(p_i + a_i) = (1-\relSave_i) c^S_i$. Further, $\sigAmount \ge 0$ denotes the relative cost savings that should be reached for a maximum number of customers. For each customer $i \in \setofCustomers$ we define a binary variable $\sigma_i \in \{0,1\}$ such that $\sigma_i = 1$ if $\relSave_i \ge \sigAmount$.

\begin{subequations}\footnotesize
	\begin{align}
	\max &\sum_{R \in \mathcal{R}} \sum_{i \in \setofCustomers_R} \sigma_i \\
	\text{s.t.}  &  \sum_{i \in \setofCustomers_R} p_i = p^R & \forall R \in \mathcal{R} \label{ip:eq}\\
	& (p_i + a_i) = (1-\relSave_i) c^S_i & \forall i \in \setofCustomers \label{ip:save}\\
	& \sigAmount \sigma_i \le \relSave_i &\forall i \in \setofCustomers \label{ip:sig}\\
	& p_i \in \mathbb{R},\quad \relSave_i \in \mathbb{R}_{\ge 0},\quad \sigma_i \in \{0,1\} & \forall i \in \setofCustomers \label{ip:defp}%\\
	\end{align}
\end{subequations}

The IP maximizes the number of customers that achieve relative cost savings of at least~$\sigAmount$. Constraints~\eqref{ip:eq} distribute the sum of fares among all customers of a pooling run. Constraints~\eqref{ip:save}\&\eqref{ip:sig} define the relative savings of all customers. The domain of the savings \eqref{ip:defp} assures that no customer is worse off than if she would have taken a solitary ride. 

Straightforwardly maximizing the number of customers with high savings causes reduced savings for customers that fall below $\sigAmount$, because the optimization will keep some customers' savings low or even at 0\% if this leads to more customers with substantial savings. To counteract this effect, we use goal programming and increase the threshold for $\sigAmount$ iteratively. For example, we can start with setting $\sigAmount^0 = 0.05$, maximizing the number of customers that save at least 5\% by participating in pooling. Let $\Sigma^0$ be the result of the optimization. In the next step, we introduce another binary variable $\sigma_i^0$ for each customer $i$ and add constraints
\begin{align*}\footnotesize
\sigAmount^0 \sigma^0_i &\le \relSave_i &\forall i \in \setofCustomers\\
\sum_i \sigma_i^0 &\ge \Sigma^0 &\forall i \in \setofCustomers,
\end{align*}
define a new threshold $\sigAmount^0 = 0.10$, and maximize the number of customers that save at least 10\% while not lowering the number of customers that save at least 5\% from the first iteration. This way, we iteratively maximize the number of customers who obtain large savings without reducing other customers' savings.
 

%% file: chapters/experiments.tex
\section{Design of Experiments}\label{sec:experiments}
We conduct a large-scale simulation to numerically evaluate the proposed matching mechanisms' effects,  using an SRO solution as a baseline. In this simulation, we use a greedy assignment mechanism to form matches. Each customer request that enters the system triggers an immediate matching decision of the \gls{abk:msp}. We implemented the simulation environment in Java and used Open Street Map\footnote{https://www.openstreetmap.org} for road network data, and OSM2PO library\footnote{https://osm2po.de} for vehicle routing decisions. 

Our experiments base on the New York Taxi and Limousine Commission data,\footnote{https://www1.nyc.gov/site/tlc/about/tlc-trip-record-data.page} which includes the origin-destination pairs of taxi trips in New York City. We focus on a representative sample of 9,388 trips, originating and ending in Manhattan, south of Central Park on August $27$
, 2015, starting between 9 am and 10 am. We calculate basic (SRO) fares with an initial charge ($\fixedPrice$) of \$2.50 plus \$0.50 per 1/5 mile\footnote{https://www1.nyc.gov/site/tlc/passengers/taxi-fare.page} ($\variablePrice{\origin_i}{\destination_i}$). We assume non-autonomous vehicles with low-wage drivers, resulting in provider cost of \$2.945 per mile \citep{BoeschBeckerEtAl2018,LanzettiSchifferEtAl2020}.

We create a customer request for each trip and consider five different value of time categories, discretizing the value of time interval  $\left[0.166 \$/\text{min},0.283 \$/\text{min}\right]$ as stated in \citet{Wadud2017} into $\{0.166\$/\text{min},\allowbreak 0.195\$/\text{min}, 0.225\$/\text{min}, 0.254\$/\text{min}, 0.283\$/\text{min} \}$ and uniformly draw a value of time $\valueofTime_i$ out of this set for each customer $i \in \setofCustomers$.
We account for a maximum customer waiting time of two, four, or six minutes and consider an \gls{abk:msp} with a fleet of 1,000--3,000 vehicles. We compare \gls{abk:pcp} and \gls{abk:ccp} matching mechanisms for this setting with different change fees, discount, and detour factors. 

We aim to exploit the impact of different matching strategies for today's ride-hailing systems with our experiments. Herein, we base our studies on analyzing the different improvement levers outlined in Section~\ref{subsec:statusQuo}. Apparently, there is neither consensus nor evidence on how customers react to the different matching strategies and how the share of customers willing to participate in ride-pooling will be affected. Hence, we extend our studies to scenarios with different matching acceptance rates (MARs), i.e., different customer share levels willing to participate in pooling. Table~\ref{tab:simparams} summarizes the parameters and treatment variables of our experiments.
\begin{table}[!hb]\singlespacing\small\centering
	\vspace{-0.4cm}
	\caption{Experiment parameters and treatment variables.}
	\centering\footnotesize\setlength{\tabcolsep}{0.12cm}
	\label{tab:simparams}
	\begin{tabular}{lll}
		\toprule
		Parameters & \gls{abk:sol} initial charge ($\fixedPrice$) & \$2.50\\
		&distance-dependent charge ($\variablePrice{\origin}{\destination}$) & \$0.50 per 1/5 mile\\
		%& provider cost & \$2.945 per mile\\
		& value of time ($\phi_c$) [$\$/\text{min}$] & $U\{0.166,0.195,0.225,0.254,0.283\}$\\
		\addlinespace
		Treatment variables &
		matching mechanism & $\{$SRO, \gls{abk:pcp}, \gls{abk:ccp}$\}$ \\
		&max. customer waiting time [seconds] & $\{120,240,360\}$\\
		&MAR [\%] & $\{10,20,\ldots,100\}$\\
		&number of vehicles &$\{1000,1500,2000,2500,3000\}$\\
		&change fee ($\changeFee$, \gls{abk:ccp} only) & $\{1.50,2.00,2.50,3.00\}$\\
		&discount factor ($\discount$, \gls{abk:pcp} only) & $\{0.70,0.75,0.80,0.85,0.90,1.00\}$\\
		&maximum detour factor ($\deviation$, \gls{abk:pcp} only) & $\{0.1,0.3,0.5\}$\\
		\bottomrule
	\end{tabular}
\end{table}

%% file: chapters/results.tex
\section{Results}\label{sec:results}
In the following, we analyze the operational impact of the studied pooling mechanisms based on our simulation. We first analyze these impacts from a system perspective (Section~\ref{subsec:systemperspective}), before we focus on an operator perspective (Section~\ref{subsec:providerperspective}) and a customer perspective (Section~\ref{subsec:userperspective}). 
\subsection{System Perspective}\label{subsec:systemperspective}
We first analyze a system's service rate, i.e., the share of customers whose requests can be served by the ride-hailing system without violating any constraints, e.g., on waiting times or detours. To this end, Table~\ref{tab:unservedCompare} shows the average share of unserved customers for a \gls{abk:sol} mechanism and for both pooling mechanism over all combinations of waiting times, number of vehicles, and mechanism settings. Even the \gls{abk:sol} mechanism yields a service rate of 99.35\% as the fleet size is sufficiently large. 

However, both pooling mechanisms allow to further decrease the number of unserved customer requests, especially for increasing MARs. Here, the \gls{abk:ccp} mechanism outperforms the \gls{abk:pcp} mechanism such that a service level of 99.97\% can be reached with the \gls{abk:ccp} mechanism for an MAR of 100\%. 
\begin{result}
	Pooling increases a ride-hailing fleet's service rate. In this context, the \gls{abk:ccp} mechanism outperforms the \gls{abk:pcp} mechanism for every MAR level.
\end{result}

Tables~\ref{tab:unservedNumTaxis} and \ref{tab:unservedWaitingTime} detail the sensitivity of this observation with respect to the fleet size (Table~\ref{tab:unservedNumTaxis}) and with respect to a customer's maximum waiting time (Table~\ref{tab:unservedWaitingTime}). Table~\ref{tab:unservedNumTaxis} shows that the impact of both pooling mechanisms on improving the customer service level increases for smaller fleet sizes. Indeed, compensating the service rate improvement of each pooling mechanism in \gls{abk:sol} operations would require a fleet size increase of 1000 vehicles. Again, the \gls{abk:ccp} mechanism outperforms the \gls{abk:pcp} mechanism in all scenarios. Table~\ref{tab:unservedWaitingTime} complements this analysis with sensitivities on the maximum customer waiting time. 
\begin{table}[!hb]
	\caption{Average percentage share of unserved customers for each mechanisms and varying MARs.\label{tab:unservedCompare}}
	\centering\footnotesize
	\begin{tabular}{lrrrrrrrrrr}
		\hline
		& \multicolumn{9}{c}{MAR [\%]}  \\
		\cmidrule(lr{1em}){2-11}
		Mechanism & 10 & 20 & 30 & 40 & 50 & 60 & 70 & 80 & 90 & 100 \\ 
		\hline
		\gls{abk:sol}  & 0.65 & 0.65 & 0.65 & 0.65 & 0.65 & 0.65 & 0.65 & 0.65 & 0.65 & 0.65 \\ 
		\gls{abk:ccp} &0.56	& 0.45	& 0.37	& 0.30	& 0.21	& 0.18	& 0.13	& 0.09	& 0.06	& 0.03 \\
		\gls{abk:pcp} & 0.60	& 0.53	& 0.43	& 0.36	& 0.28	& 0.26	& 0.19	& 0.15	& 0.12	& 0.09 \\
		\hline
	\end{tabular}
\end{table}
Both pooling mechanisms 
\begin{table}[!hb]
	\centering\footnotesize
	\begin{minipage}{0.48\textwidth}
		\caption{Average percentage share of unserved customers for each mechanism and varying fleet size.\label{tab:unservedNumTaxis}} 
		\centering
		\begin{tabular}{lrrrrr}
			\hline
			& \multicolumn{5}{c}{number of vehicles}  \\
			\cmidrule(lr{1em}){2-6}
			Mechanism  & 1000 & 1500 & 2000 & 2500 & 3000 \\ 
			\hline
			\gls{abk:sol} & 1.91 & 0.74 & 0.33 & 0.16 & 0.11 \\ 
			\gls{abk:ccp} & 0.77 & 0.24	& 0.09 & 0.05	& 0.03 \\
			\gls{abk:pcp} & 0.97 & 0.31	& 0.12 & 0.06	& 0.04 \\
			\hline
		\end{tabular}
	\end{minipage}
	\hfill
	\begin{minipage}{0.48\textwidth}
		\caption{Average percentage share of unserved customers for varying maximal customer waiting times.\label{tab:unservedWaitingTime}}
		\centering
		\begin{tabular}{lrrr}
			\hline
			& \multicolumn{3}{c}{max. waiting time}  \\
			\cmidrule(lr{1em}){2-4}
			Mechanism  & 120 & 240 & 360 \\ 
			\hline
			\gls{abk:sol} & 0.99 & 0.12 & 0.0021 \\ 
			\gls{abk:ccp} & 0.69 & 0.02 & 0.0009 \\ 
			\gls{abk:pcp} & 0.86 & 0.04 & 0.0010 \\ 
			\hline
		\end{tabular}
	\end{minipage}
\end{table}
outperform the SRO mechanism, and the \gls{abk:ccp} mechanism outperforms the \gls{abk:pcp} mechanism, in particular if the maximum customer waiting time is small.
\begin{result}
	The impact of pooling on the fleet's service level compared to SRO is higher for smaller fleet sizes and low customer waiting times. In any case, both pooling mechanisms perform better than solitary operations, and the \gls{abk:ccp} mechanism performs better than the \gls{abk:pcp} mechanism.
\end{result}

From this initial analysis, we conclude that a fleet of 2,000 vehicles is sufficient to serve around 99.9\% of trips in our studies. As we are interested in the difference between \gls{abk:pcp} and \gls{abk:ccp} pooling mechanisms rather than in the general savings potential of pooling in undersized fleets, we thus fix the fleet size to 2,000 vehicles for all subsequent analysis.

\begin{figure}[!hb]
	\centering
	\begin{center}
		\includegraphics[width=\textwidth]{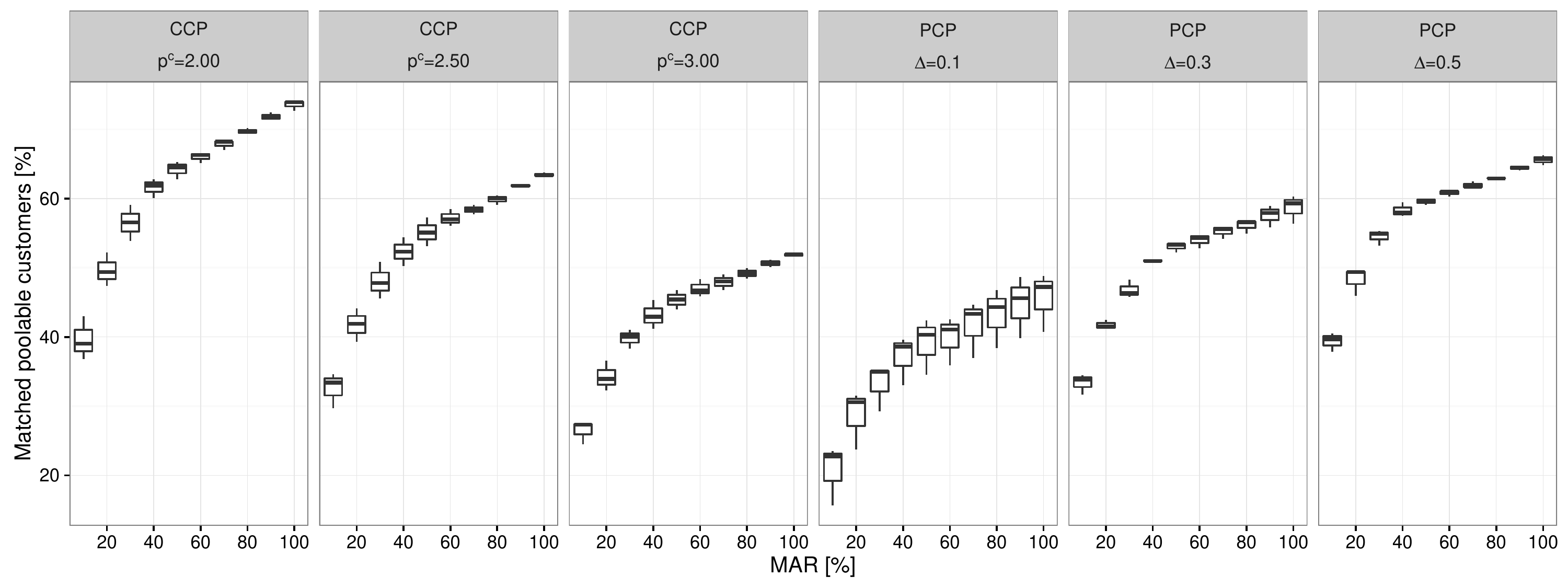}
	\end{center}
	\caption{Percentage of pooled customers for different mechanisms and MARs, as a share of the population of poolable customers.\label{fig:relPoolPlot}}
\end{figure}

Next, we analyze the share of pooled customers for each pooling mechanism. Figure~\ref{fig:relPoolPlot} shows the percentage of pooled customers over varying MARs for different \gls{abk:pcp} and \gls{abk:ccp} settings, based on the population of poolable customers.\footnote{Trivially, the share of pooled customers as part of the overall population increases with increasing MARs.} Both pooling mechanisms are susceptible to the respective MAR. While we observe large increases in the number of pooled customers for MAR increases up to 40\%, we observe diminishing improvements for higher MARs. Additionally, the \gls{abk:ccp} mechanism is sensitive to its change fee such that a 50\% increase in $p^c$ may cause a decrease in the share of pooled customers of up to 20 percentage points for high MARs. The \gls{abk:pcp} mechanism remains sensitive to the maximum detour threshold such that one may increase the share of pooled customers up to 20 percentage points by a 40 percentage point increase in $\Delta$. Moreover, the PCP mechanism yields less stable results for lower $\Delta$, while the \gls{abk:ccp} mechanisms show in general smaller deviations in between different simulation runs. 
\begin{result}
	The number of pooled customers grows with the population of poolable customers. The relative growth is substantial for MARs up to 40\%.
\end{result}

\begin{figure}[!hb]
	\centering
	\begin{center}
		\includegraphics[width=\textwidth]{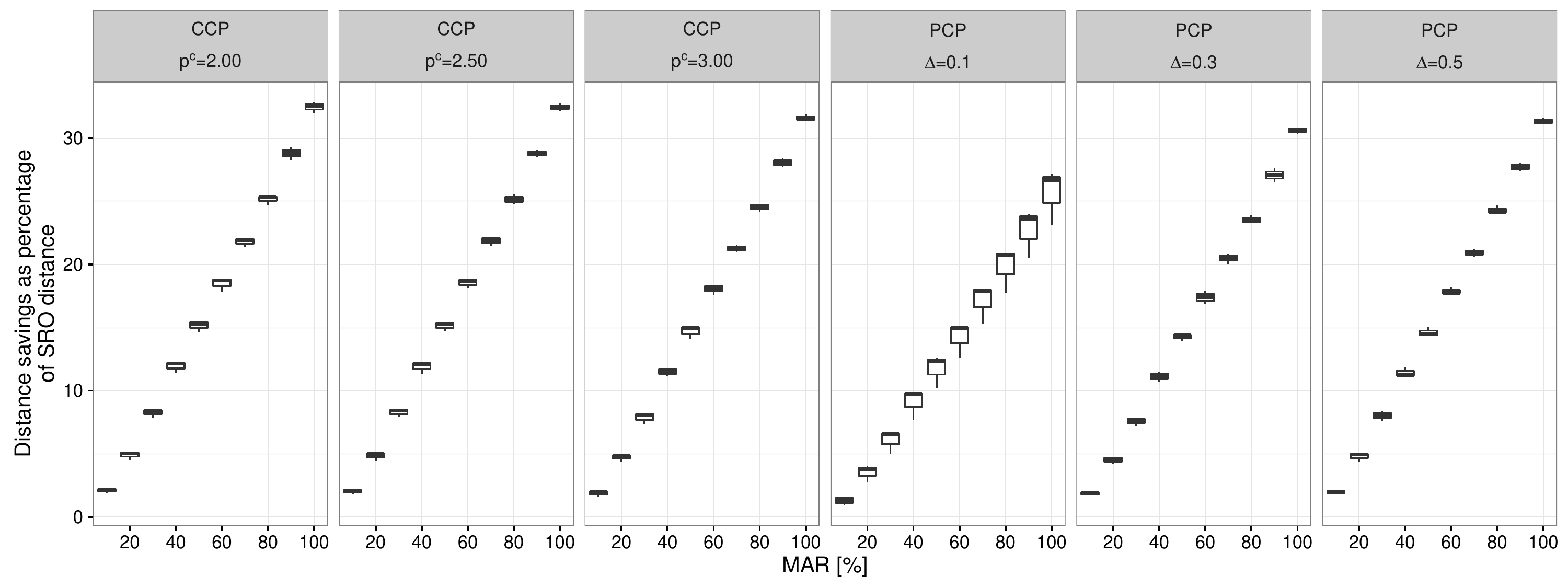}
		{\footnotesize
			\begin{tabular}{lrrrrrrrrrrrr}\\
				\hline
				& & & & \multicolumn{9}{c}{MAR}  \\
				\cmidrule(lr{1em}){4-12}
				Mechanism & $\changeFee$ & $\deviation$ & 10 & 20 & 30 & 40 & 50 & 60 & 70 & 80 & 90 & 100 \\ 
				\hline
				\gls{abk:ccp} & 2.00 && 2.08	&	4.90	&	8.24	&	11.92	&	15.16	&	18.47	&	21.8	&	25.15	&	28.81	&	32.48 \\ 
				\gls{abk:ccp} & 2.50 && 2.03	&	4.83	&	8.25	&	11.91	&	15.12	&	18.55	&	21.86	&	25.16	&	28.78	&	32.46 \\ 
				\gls{abk:ccp} & 3.00 && 1.86	&	4.71	&	7.86	&	11.50	&	14.70	&	18.05	&	21.26	&	24.53	&	28.05	&	31.63 \\ 
				\gls{abk:pcp}  & & 0.1 & 1.29	&	3.52	&	6.04	&	9.08	&	11.72	&	14.17	&	17.07	&	19.78	&	22.7	&	25.66 \\ 
				\gls{abk:pcp}  & & 0.3 & 1.84	&	4.51	&	7.56	&	11.12	&	14.26	&	17.38	&	20.48	&	23.55	&	27.08	&	30.62 \\ 
				\gls{abk:pcp}  & & 0.5  & 1.94	&	4.77	&	8.02	&	11.44	&	14.61	&	17.88	&	20.92	&	24.31	&	27.73	&	31.35 \\ 
				\hline
		\end{tabular}}
	\end{center}
	\caption{Average distance savings for different mechanisms and MARs as percentage of the distance when employing \gls{abk:sol}.}\label{fig:distanceSavings}
\end{figure}

We focus on reducing a fleet's overall traveled distance to analyze potential emission or energy savings, which allows us to quantify respective savings without specifying a fleet's drivetrain technology. Figure~\ref{fig:distanceSavings} shows the fleet's average traveled distance \textit{savings} for the \gls{abk:ccp} and the \gls{abk:pcp} mechanism and different values of $p^c$ and~$\Delta$. To this end, we note that the discount factor $\delta$ does not affect pooling decisions in the \gls{abk:pcp} mechanism as customers receive the discount independent of the pooling decision. Accordingly, we exclude $\delta$ from the following analysis. For MARs above 40\%, both mechanisms allow for substantial distance savings of more than 9\% and up to 32\%. For the studied parameter ranges, the \gls{abk:ccp} mechanism always yields higher distance savings, particularly for high MARs where we observe differences of up to 7 percentage points.
\begin{result}
	Independent of the pooling mechanism, pooling allows for distance savings of up to 32\%. Herein, the \gls{abk:ccp} mechanism yields better results than the \gls{abk:pcp} mechanism, in particular for high MARs.
\end{result}

In our scenario, a 5\% reduction in distance equals a reduction of 0.16 metric tons of carbon dioxide emissions.\footnote{\url{https://www.eea.europa.eu/ds_resolveuid/93740c18c4a4489faf10bb84a462e438}}
Noting that these emission savings relate to a single 30 minute time interval, our results emphasize the significant potential that passenger pooling bears from an environmental perspective, even for relatively small MARs.

\subsection{Provider Perspective}\label{subsec:providerperspective}
\begin{figure}[!hb]
	\centering\singlespacing\footnotesize
	\centering
	\begin{subfigure}{0.49\textwidth}
		\includegraphics[width=\textwidth]{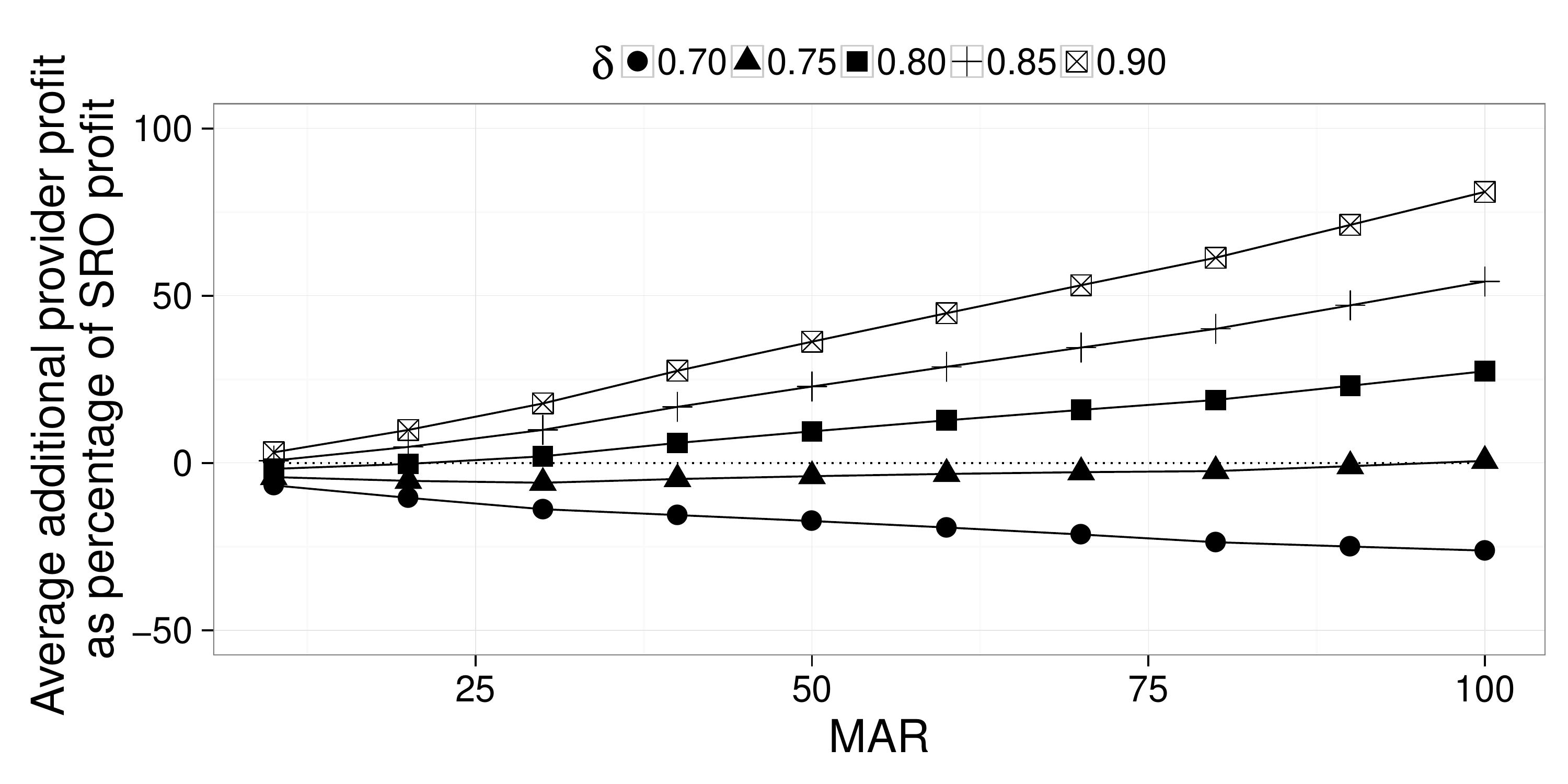}
		\caption{PCP profit for varying $\delta$ with $\Delta = 0.3$.}\label{subfig:edgeProfitsPCP}
	\end{subfigure}
	\begin{subfigure}{0.49\textwidth}
		\includegraphics[width=\textwidth]{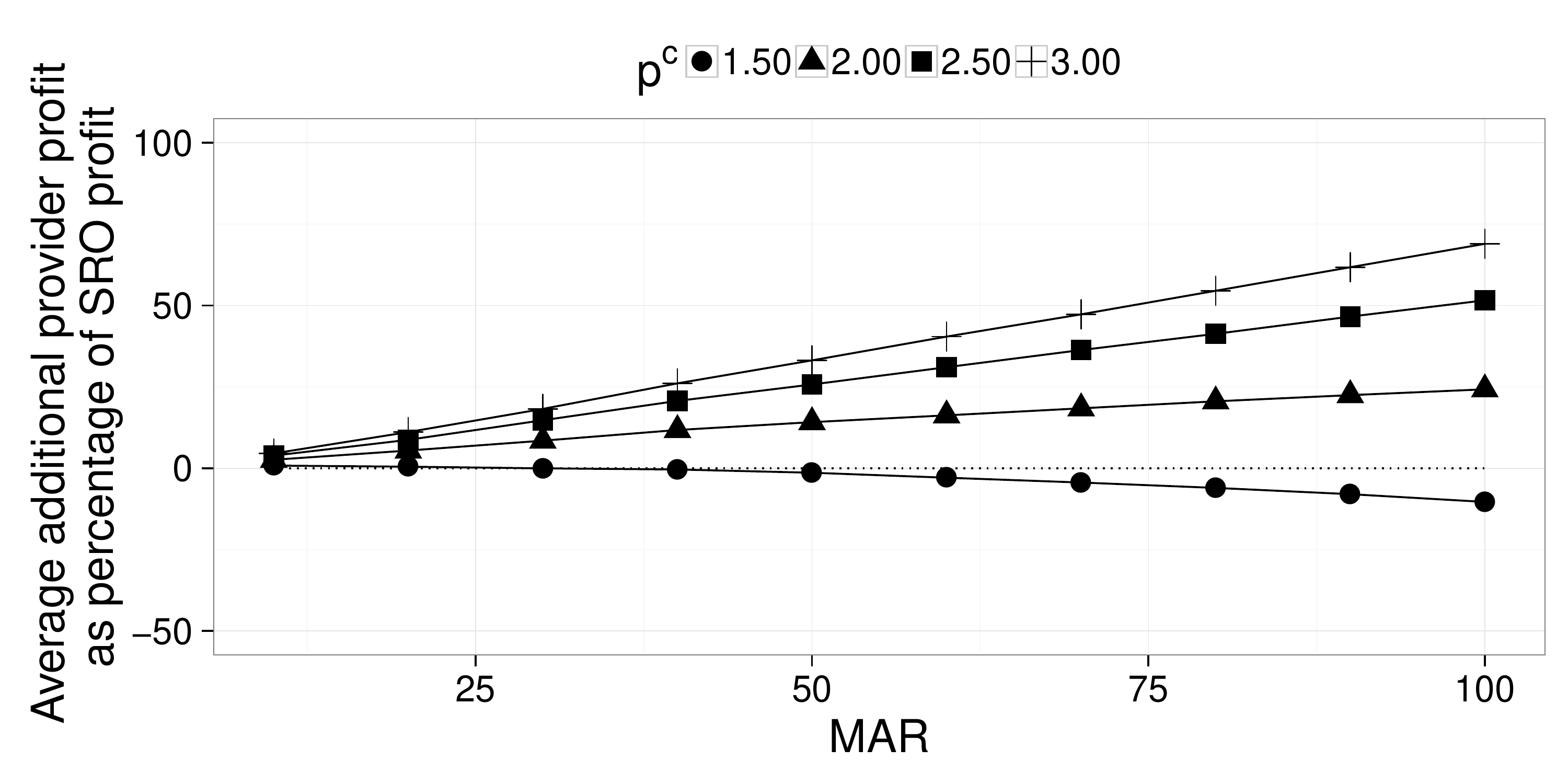}
		\caption{CCP profit for varying $p^c$.}\label{subfig:edgeProfitsCCP}
	\end{subfigure}
	\caption{Additional profit of \gls{abk:pcp} and \gls{abk:ccp} compared to SRO for varying $\delta$ and $p^c$.}\label{fig:edgeProfits}
	\vspace{-0.5cm}
\end{figure}

We now analyze the difference between both pooling mechanisms from a ride-hailing provider's perspective, i.e., we analyze the impact of both mechanisms on a fleet's profit. Naturally, an operator will only use a pooling mechanism if it yields increased profits compared to operating solitary rides. Therefore, we first analyze for which parameters each mechanism is viable. Figure~\ref{fig:edgeProfits} shows the profit for both pooling mechanisms over varying MARs, compared to the profits for solitary operation. We detail the \gls{abk:pcp} profits for varying discount factors $\delta\in\{0.7,0.75,0.8,0.85,0.9\}$ in Figure~\ref{subfig:edgeProfitsPCP}, while Figure~\ref{subfig:edgeProfitsCCP} shows the \gls{abk:ccp} profits for varying change fees $p^c\in\{1.5,2.0,2.5,3.0\}$. For the sake of conciseness, Figure~\ref{subfig:edgeProfitsPCP} only shows results for a detour factor of $\Delta = 0.3$ as we observe similar trends for other detour factors.

Notably, both mechanisms only show increasing profit curves for a change fee of at minimum $p^c \ge 2.00$ (\gls{abk:ccp}) or a discount factor of at most $\discount=0.8$ (\gls{abk:pcp}). Within these settings, the provider may strive for a larger MAR in order to maximize its profits. Remarkably, we observe an adverse behavior for both pooling mechanisms with $\delta = 0.7$ or $p^c = 1.5$. In these cases, pooled operations yield less profits than \gls{abk:sol} operations, with increasing losses for increasing MARs. Here, each additional poolable customer leads to a profit decrease, because either her guaranteed discount outweighs the cost reduction from pooling her (\gls{abk:pcp}), or the low change fee leads to poolings being beneficial for customers, but not for the provider (\gls{abk:ccp}). The latter effect shows that the profitability of a \gls{abk:ccp} matching algorithm still depends on the right parameter setting, although it avoids sunk costs in unsuccessful pooling by design (cf. Section~\ref{subsec:properties}).
\begin{result}
	Provider profits increase with increasing matching acceptance rates, except for \gls{abk:ccp} mechanisms with low change fees $p^c$ or \gls{abk:pcp} mechanisms with low $\delta$.
\end{result}
Based on these initial results, we limit subsequent analyses to $\delta \ge 0.8$ and $p^c \ge 2.00$.

Table \ref{tab:profit_main} further details for various settings of $p^c$, $\Delta$, and $\delta$, and varying MARs, the average additional profit for \gls{abk:pcp} and \gls{abk:ccp} mechanisms as a percentage of the profit that can be achieved in an \gls{abk:sol} setting. Both pooling mechanisms may yield substantial profit increases, in particular for MARs of 50\% or higher. However, the viability of both pooling mechanisms may vary significantly dependent on the respective parameter settings. The \gls{abk:pcp} mechanism may yield decreasing profits for $\delta = 0.8$ if $\Delta$ and the MAR is small. While the \gls{abk:pcp} mechanism yields 
higher profit gains 
\begin{table}[!ht]
	\caption{Average additional profit for varying MARs compared to profits under \gls{abk:sol} (in percent).\label{tab:profit_main}}
	\centering\footnotesize\setlength{\tabcolsep}{0.15cm}
	\begin{tabular}{lrrrrrrrrrrrrr}
		\hline
		& & & & \multicolumn{9}{c}{MAR}  \\
		\cmidrule(lr{1em}){5-14}
		Mechanism & $\changeFee$ & $\discount$ & $\deviation$ & 10 & 20 & 30 & 40 & 50 & 60 & 70 & 80 & 90 & 100 \\ 
		\hline		
		\gls{abk:ccp} & 2.00 & & & 2.59 & 5.42 & 8.46 & 11.74 & 14.15 & 16.24 & 18.40 & 20.55 & 22.46 & 24.24 \\ 
		\gls{abk:ccp} & 2.50 & & & 3.86 & 8.72 & 14.74 & 20.68 & 25.73 & 31.03 & 36.32 & 41.29 & 46.59 & 51.59 \\ 
		\gls{abk:ccp} & 3.00 & & & 4.53 & 11.11 & 18.22 & 26.07 & 33.14 & 40.44 & 47.29 & 54.53 & 61.76 & 68.96 \\ 
		\gls{abk:pcp} &	&0.8	&0.1	&-4.12	&-4.59	&-4.64	&-2.62	&-1.60	&-1.28	&0.97	&2.28	&4.01	&5.78\\ 
		\gls{abk:pcp} &	&0.8	&0.3	&-1.77	&-0.29	&1.97	&5.97	&9.44	&12.73	&15.86	&18.82	&23.07	&27.41\\ 
		\gls{abk:pcp} &	&0.8	&0.5	&-1.25	&0.89	&4.03	&7.43	&11.07	&14.99	&17.81	&22.10	&25.85	&30.60\\ 
		\gls{abk:pcp} &	&0.85	&0.1	&-1.65	&0.48	&3.26	&8.17	&11.80	&14.73	&19.60	&23.54	&28.04	&32.57\\ 
		\gls{abk:pcp} &	&0.85	&0.3	&0.69	&4.79	&9.87	&16.76	&22.84	&28.75	&34.49	&40.07	&47.11	&54.22\\ 
		\gls{abk:pcp} &	&0.85	&0.5	&1.22	&5.97	&11.93	&18.22	&24.48	&31.00	&36.44	&43.36	&49.89	&57.41\\ 
		\gls{abk:pcp} &	&0.9	&0.1	&0.82	&5.55	&11.17	&18.95	&25.21	&30.75	&38.23	&44.79	&52.08	&59.37\\ 
		\gls{abk:pcp} &	&0.9	&0.3	&3.16	&9.86	&17.77	&27.54	&36.24	&44.76	&53.12	&61.33	&71.14	&81.03\\ 
		\gls{abk:pcp} &	&0.9	&0.5	&3.69	&11.05	&19.84	&29.00	&37.89	&47.02	&55.07	&64.61	&73.92	&84.21\\ 
		\hline
	\end{tabular}	
\end{table}
for favorable parameter settings, it also tends to show a higher variation between different settings; in particular, it has more settings on which the profit decreases. On the contrary, the \gls{abk:ccp} mechanism shows a lower maximum profit increase but more stable results with respect to potential losses. Here, it is worth noting that the \gls{abk:pcp} mechanism outperforms the \gls{abk:ccp} mechanism only for settings in which customers accept a small discount and a large detour. Accordingly, it remains questionable how many customers may participate in such a pooling system in practice. 
\begin{result}
	Both pooling mechanisms allow for large profit increases but remain sensitive to their respective parameter settings. While the \gls{abk:pcp} mechanism may allow for the largest profit increase, the \gls{abk:ccp} mechanism remains more robust with respect to potential losses.
\end{result}

A higher detour factor increases an operator's operational options, such that one would expect a proportional improvement in the \gls{abk:msp}'s returns. This effect holds true when increasing $\Delta$ from 10\% to 30\%, regardless of the MAR, but falls short for an increase from 30\% to 50\%. For example, for $\discount = 0.85$, the provider's profit does not differ significantly between $\deviation= 0.3$ and $\deviation = 0.5$ and a corresponding Wilcoxon signed-rank test yields $p = 0.7766$ such that we observe diminishing returns. This effect results from the high availability of vehicles: In many cases, using a nearby empty vehicle to serve a customer alone is more efficient than making a large detour.
\begin{result}
	Increasing the detour factor $\Delta$ in a \gls{abk:pcp} mechanism may yield diminishing returns for the provider if the fleet size is sufficiently large.
\end{result}

\subsection{Customer Perspective}\label{subsec:userperspective}
In the following, we compare the different pooling mechanisms concerning their impact on customer costs. Although higher discounts, in general, imply lower total costs for customers, we keep the limited parameter selection from Section~\ref{subsec:providerperspective} to account for an operator's rational constraints. To analyze cost effects, one may use three different customer populations: all customers in the system, poolable customers, and finally pooled customers. In the following, we compare poolable customers' costs because it allows analyzing the benefits for customers to set themselves poolable directly. 

\begin{figure}[!hb]
	\centering\singlespacing\footnotesize
	\centering
	\begin{subfigure}{0.65\textwidth}
		\includegraphics[width=\textwidth]{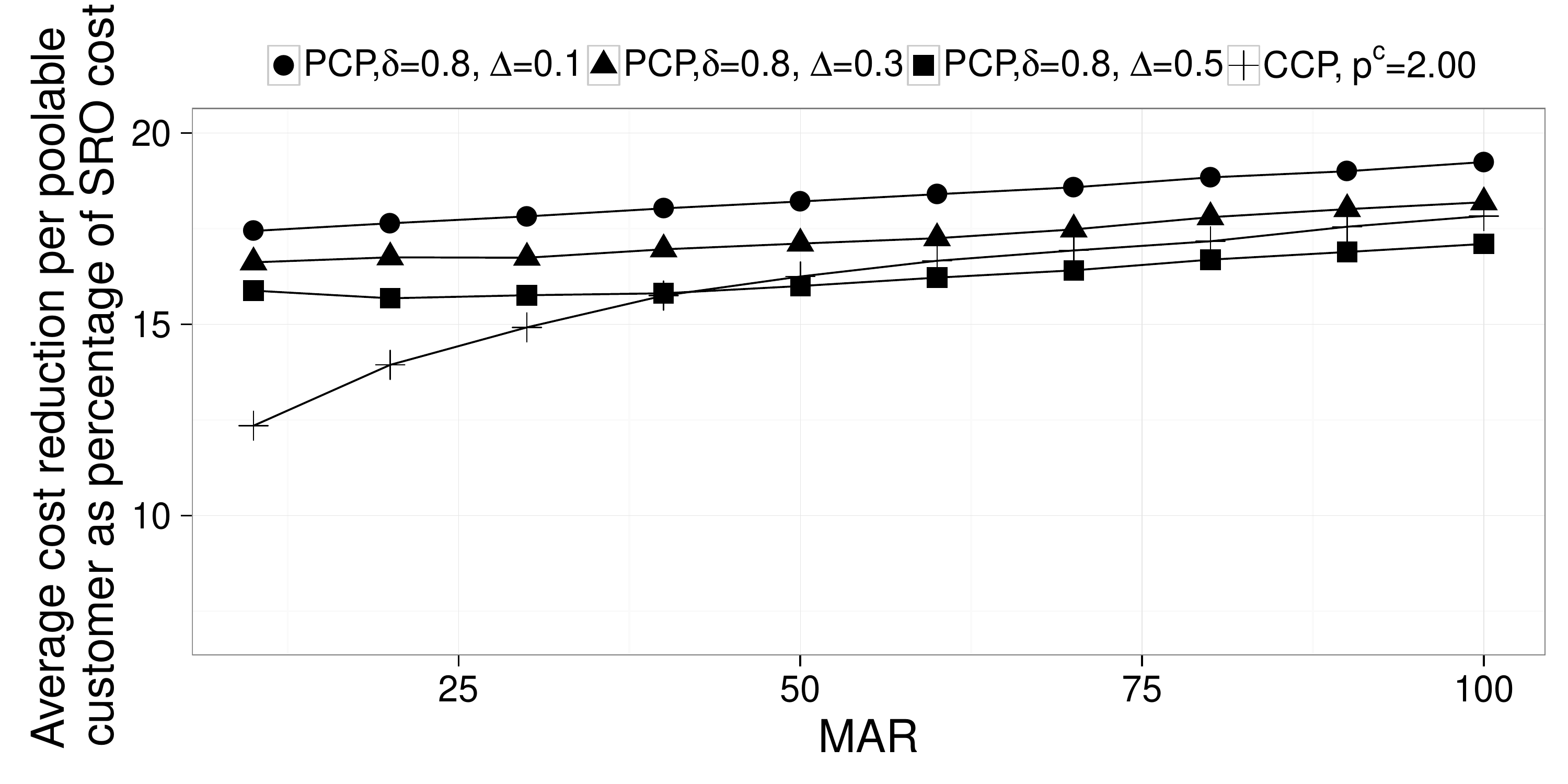}
	\end{subfigure}
	\begin{subfigure}{0.65\textwidth}
		\includegraphics[width=\textwidth]{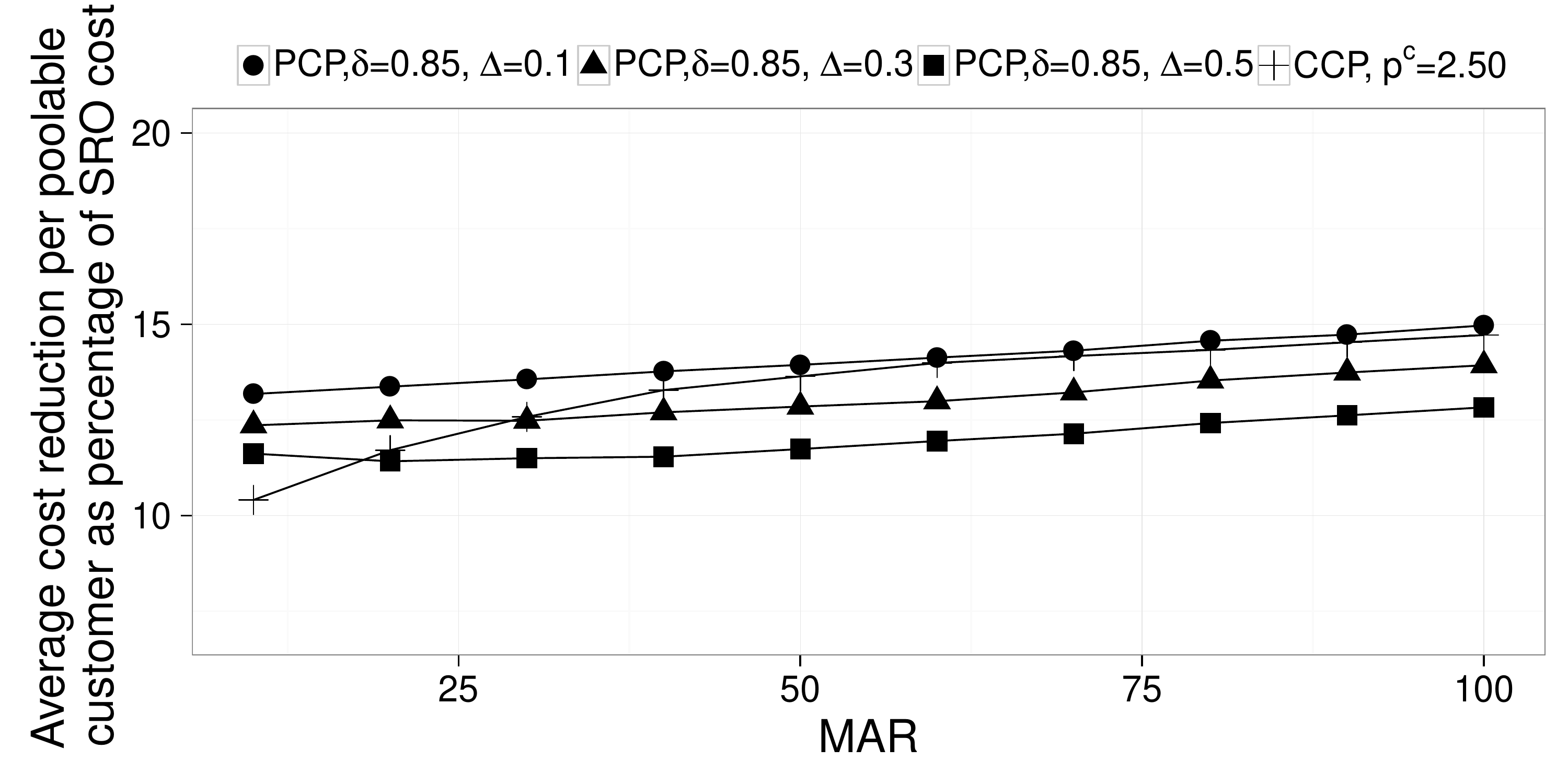}
	\end{subfigure}
	\begin{subfigure}{0.65\textwidth}
		\includegraphics[width=\textwidth]{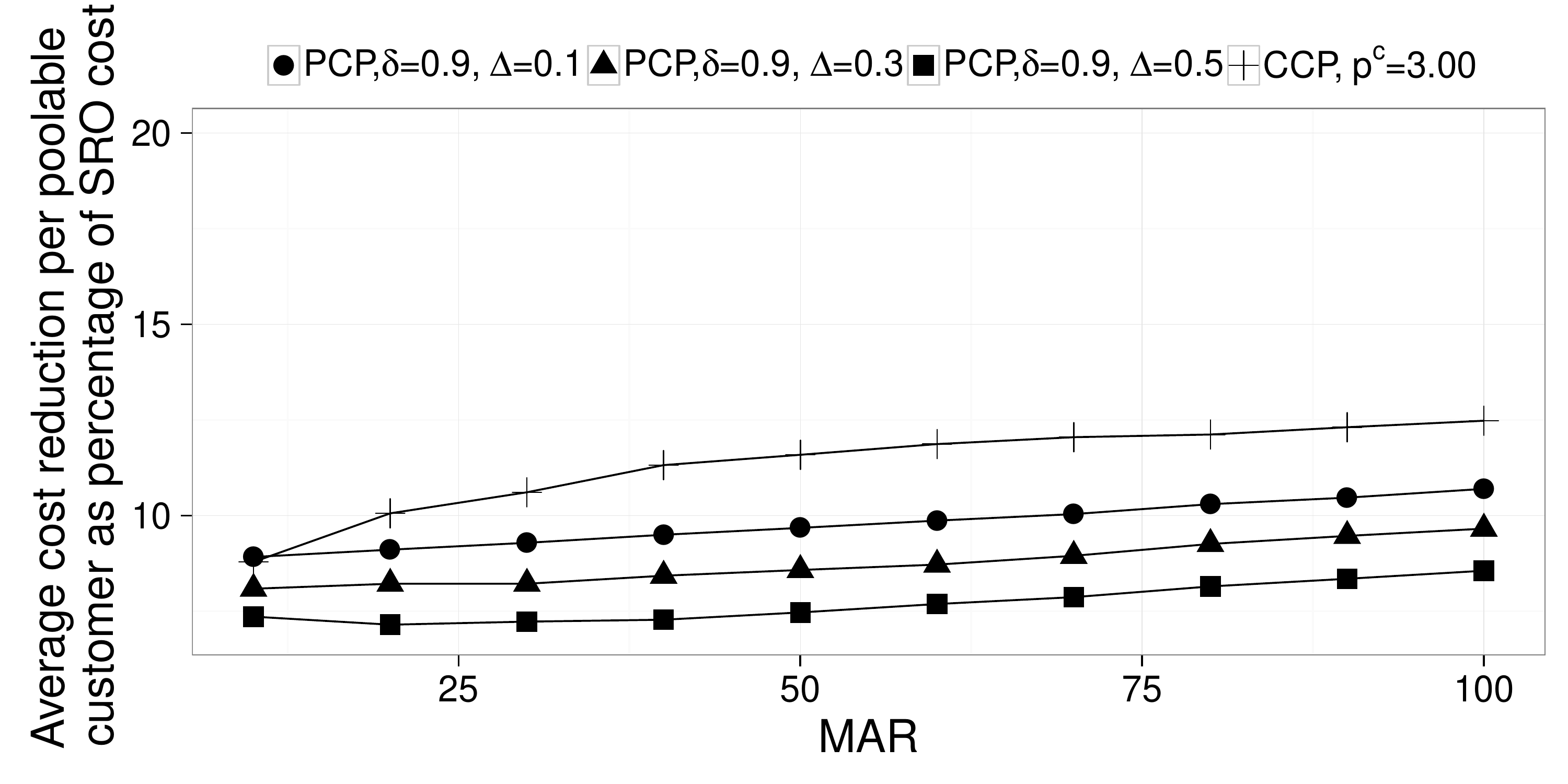}
	\end{subfigure}
	\caption{Average cost reduction per poolable customer for different mechanisms and MARs as percentage of \gls{abk:sol} cost.\label{fig:costsAll}}
\end{figure}

Figure~\ref{fig:costsAll} shows the decrease in cost per poolable customer as percentage share of the solitary ride costs over all MARs. To improve readability we group the pooling mechanisms with regard to their similarity in costs and profits. Table~\ref{tab:costs_poolable} complements our cost analysis by showing the average cost per poolable customer over varying MARs for different \gls{abk:pcp} and \gls{abk:ccp} settings. 

Analogously to positive system effects and provider profits, average cost reductions increase with higher MARs. When using a \gls{abk:ccp} mechanism, customers gain on average a minimum cost savings of about 9\% independent of the respective MAR and for any $p^c$. With higher MARs, this savings may increase up to 18\%. Potential cost savings for the \gls{abk:pcp} mechanism tend to have a larger variance, dependent on the respective parameter settings.
\begin{table}[!ht]
	\caption{Average cost reduction per poolable customer for different mechanisms and MARs as compared to \gls{abk:sol} (in percent).\label{tab:costs_poolable}}
	\centering\footnotesize\setlength{\tabcolsep}{0.15cm}
	\begin{tabular}{lrrrrrrrrrrrrr}
		\hline
		& & & & \multicolumn{9}{c}{MAR}  \\
		\cmidrule(lr{1em}){5-14}
		Mechanism & $\changeFee$ & $\discount$ & $\deviation$ & 10 & 20 & 30 & 40 & 50 & 60 & 70 & 80 & 90 & 100 \\ 
		\hline
		\gls{abk:ccp}		& 2.00&	& & 12.35 & 13.94 & 14.92 & 15.75 & 16.25 & 16.66 & 16.93 & 17.17 & 17.55 & 17.83 \\
		\gls{abk:pcp} &	& 0.80	& 0.1	& 17.44 & 17.64 & 17.82 & 18.03 & 18.21 & 18.40 & 18.58 & 18.84 & 19.00 & 19.24 \\
		\gls{abk:pcp} &	& 0.80	& 0.3	& 16.62 & 16.75 & 16.74 & 16.96 & 17.11 & 17.25 & 17.48 & 17.80 & 18.01 & 18.19\\
		\gls{abk:pcp} &	& 0.80	& 0.5	& 15.88 & 15.68 & 15.76 & 15.81 & 16.00 & 16.22 & 16.41 & 16.69 & 16.89 & 17.10\\
		\hline
		\gls{abk:ccp}		& 2.50&	& & 10.41 & 11.71 & 12.58 & 13.28 & 13.64 & 13.99 & 14.17 & 14.33 & 14.53 & 14.72\\
		\gls{abk:pcp} &	& 0.85	& 0.1& 13.18 & 13.37 & 13.56 & 13.77 & 13.94 & 14.13 & 14.31 & 14.57 & 14.73 & 14.97 \\
		\gls{abk:pcp} &	& 0.85	& 0.3	& 12.36 & 12.49 & 12.48 & 12.70 & 12.85 & 12.99 & 13.22 & 13.53 & 13.74 & 13.93  \\
		\gls{abk:pcp} &	& 0.85	& 0.5	& 11.62 & 11.42 & 11.50 & 11.54 & 11.74 & 11.95 & 12.14 & 12.42 & 12.62 & 12.83  \\
		\hline
		\gls{abk:ccp}		& 3.00&	& & 8.79 & 10.06 & 10.61 & 11.32 & 11.59 & 11.87 & 12.05 & 12.12 & 12.31 & 12.48  \\
		\gls{abk:pcp} &	& 0.90	& 0.1	& 8.92 & 9.11 & 9.29 & 9.50 & 9.68 & 9.87 & 10.04 & 10.30 & 10.47 & 10.70 \\
		\gls{abk:pcp} &	& 0.90	& 0.3	& 8.09 & 8.22 & 8.22 & 8.43 & 8.58 & 8.72 & 8.95 & 9.26 & 9.47 & 9.66 \\
		\gls{abk:pcp} &	& 0.90	& 0.5	& 7.36 & 7.15 & 7.23 & 7.28 & 7.47 & 7.69 & 7.87 & 8.15 & 8.35 & 8.56\\
		\hline
	\end{tabular}
\end{table}

Remarkably, even for a \gls{abk:pcp} mechanism without any discount ($\delta=1.00$), poolable customers may experience a slight cost reduction for $\Delta = 0.1$. When considering the total population of customers, a \gls{abk:pcp} mechanism with $\delta=1.00$ leads to cost savings even for larger $\Delta$. This effect results because pooling allows for a more efficient usage of vehicles and hence for earlier customer arrivals, which reduces a customer's total cost without any discount on customer payments.
\begin{result}
	Pooling is cost-beneficial for customers independent of the pooling mechanism. For the \gls{abk:pcp} mechanism, slight cost reductions exist even without a discount, especially for high MARs.
\end{result}

One advantage of the \gls{abk:ccp} mechanism is its flexible pricing, which allows to design cost sharing among customers such that as many customers as possible save a substantial cost share compared to a solitary ride (cf. Section~\ref{sec:flexPricing}). In the following, we detail such cost shares focusing on two different pricing schemes for the \gls{abk:ccp} mechanism, which we refer to as \gls{abk:ccp}(Shapley) and \gls{abk:ccp}(IP), and various \gls{abk:pcp} settings. In \gls{abk:ccp}(Shapley), the savings for two pooled customers are shared according to the Shapley value. For \gls{abk:ccp}(IP), we employ the IP with goal programming as described in Section \ref{sec:flexPricing}. We first maximize the number of customers that save at least 5\% of costs as compared to a solitary ride and then iteratively increase the threshold to maximize the number of customers saving at least 10\%, 15\%, and 20\% of costs.

Table \ref{tab:significant_percentage} shows the customers' share that saves a minimum percentage share of costs for the described mechanisms for MARs of 20\% and 50\%. For the sake of conciseness, we only report results for these two settings, which represent a thinner and a thicker market of poolable customers. The tendency of the reported results holds true for other MAR values. We note that no customer makes a loss by design when participating in a \gls{abk:ccp} mechanism. For the \gls{abk:pcp} settings, corner cases exist in which some customers are worse off compared to not participating in the pooling mechanism. For example, given a $\delta$ of $0.9$, there exist customers that would have received a ride with a lower total cost if they would not have set themselves poolable.

In the low-cost setting, the \gls{abk:pcp} mechanisms achieve higher customer shares with savings up to 15\%, 
\begin{table}[!ht]
	\caption{Overview of customers' savings when setting themself poolable for \gls{abk:pcp} mechanisms and \gls{abk:ccp} mechanisms (in percentage of poolable customers, compared to not setting themselves poolable). \label{tab:significant_percentage}}
	\centering\footnotesize\setlength{\tabcolsep}{0.06cm}
	\begin{tabular}{lrrrrrrrrrrrrr}
		\hline
		& & & & \multicolumn{5}{c}{MAR = 20\%, total cost improves} & \multicolumn{5}{c}{MAR = 50\%, total cost improves} \\
		\cmidrule(lr{1em}){5-9}\cmidrule(lr{1em}){10-14}
		Mechanism & $\changeFee$ & $\discount$ & $\deviation$ & $\ge 0\%$ & $\ge 5\%$ & $\ge 10\%$ & $\ge 15\%$ & $\ge 20\%$ & $\ge 0\%$ & $\ge 5\%$ & $\ge 10\%$ & $\ge 15\%$ & $\ge 20\%$ \\ 
		\hline
		\gls{abk:ccp}(Shapley)	&	 2.00&&	&  100.00&	78.13&	52.00&	30.05&	13.55&	100.00&	80.64&	56.72&	33.22&	15.94 \\
		\gls{abk:ccp}(IP)	&	 2.00&&	&  100.00& 	91.44& 	70.40& 	41.82& 	18.15 & 100.00	& 94.52& 	77.45& 	36.11& 	31.39 \\
		\gls{abk:pcp}	&& 0.80	& 0.1	& 100.00	& 99.96	& 99.89	& 89.65	& 14.21	& 100.00	& 99.99	& 99.88	& 92.41	& 13.87 \\
		\gls{abk:pcp}	&& 0.80	& 0.3	& 100.00	& 100.00	& 98.48	& 69.31	& 10.16	& 100.00	& 99.97	& 98.68	& 70.31	& 9.34 \\
		\gls{abk:pcp}	&& 0.80	& 0.5	& 100.00	& 99.62	& 93.79	& 54.14	& 7.93	& 100.00	& 99.75	& 94.56	& 52.84	& 6.81 \\ \hline
		\gls{abk:ccp}(Shapley)	&	 2.50&&	&  100.00&	69.65&	41.59&	20.52&	7.28 &	100.00&	72.09&	43.26&	21.00&	8.53 \\
		
		\gls{abk:ccp}(IP)	&	 2.50&&	&  100.00& 	84.03& 	57.66& 	28.61& 	10.97 & 100.00	& 88.37& 	61.85& 	23.08& 	18.29 \\
		\gls{abk:pcp}	&& 0.85	& 0.1	& 99.96	& 99.89	& 96.92	& 17.94	& 4.51	& 100.00	& 99.92	& 97.58	& 18.04	& 4.26 \\
		\gls{abk:pcp}	&& 0.85	& 0.3	& 100.00	& 99.47	& 79.29	& 12.84	& 3.20	& 99.98	& 99.47	& 80.06	& 11.80	& 3.00 \\
		\gls{abk:pcp}	&& 0.85	& 0.5	& 99.88	& 96.33	& 61.77	& 10.12	& 2.33	& 99.91	& 96.72	& 60.88	& 8.74	& 2.17 \\
		\hline
		\gls{abk:ccp}(Shapley)	&	 3.00&&	&  100.00&	62.41&	32.43&	13.92&	5.48 &	100.00&	63.42&	33.15&	14.31&	5.46 \\
		\gls{abk:ccp}(IP)	&	 3.00&&	&  100.00& 	77.99& 	46.94& 	21.35& 	8.56  & 100.00	& 80.06& 	48.20& 	16.48& 	12.62 \\
		\gls{abk:pcp}	&& 0.90	& 0.1	& 99.89	& 99.46	& 23.45	& 6.02	& 1.25	& 99.94	& 99.54	& 23.38	& 5.58	& 1.25 \\
		\gls{abk:pcp}	&& 0.90	& 0.3	& 99.89	& 87.00	& 16.62	& 4.31	& 1.13	& 99.84	& 88.02	& 14.80	& 3.89	& 1.03 \\
		\gls{abk:pcp}	&& 0.90	& 0.5	& 97.98	& 69.35	& 12.99	& 3.34	& 0.80	& 98.14	& 69.44	& 11.23	& 2.79	& 0.69 \\
		\hline
	\end{tabular}
\end{table}
while the \gls{abk:ccp} mechanism achieves a higher customer share with savings of at minimum 20\%. For medium-cost and high-cost settings, the \gls{abk:ccp} mechanisms start to outperform the \gls{abk:pcp} mechanisms at cost savings of 15\% and 10\% respectively. This effect shows that the \gls{abk:pcp} mechanism is strongly constrained by its fixed discount, whereas the \gls{abk:ccp} mechanism can handle customer specific discounts more flexible, especially for higher cost-settings.

Comparing the \gls{abk:ccp} pricing schemes, we note that employing the goal programming approach increases the number of customers with high cost savings of more than 15\%, without worsening the share of customers that save at minimum 5\% or 10\% of costs. Accordingly, the goal programming approach outperforms the Shapley-based pricing.
\begin{result}
	The number of customers that receive substantial savings by setting themselves poolable is larger for \gls{abk:ccp} than for \gls{abk:pcp} mechanisms. For \gls{abk:pcp} mechanisms with a discount factor of $\delta = 0.9$, some pooled customers may have a higher total cost than for a solitary ride.
\end{result}

%% file: chapters/winwin.tex
\section{Synthesis}\label{sec:synthesis}
In the following, we summarize the trade-offs and synergy potential, which we observed in Section~\ref{sec:results}. For this discussion, we define three categories of matching mechanisms: \textit{low-cost mechanisms} are a \gls{abk:ccp} mechanism with $p^C=2.0$ and a \gls{abk:pcp} mechanism with $\delta = 0.8$; we refer to a \gls{abk:ccp} mechanism with $p^C=2.5$ and a \gls{abk:pcp} mechanism with $\delta = 0.85$ as \textit{medium-cost mechanisms}; and to a \gls{abk:ccp} mechanism with $p^C=3.0$ and a \gls{abk:pcp} mechanism with $\delta = 0.9$ as \textit{high-cost mechanisms}.

In general, we observe that whenever a pooling mechanism is profitable for the provider, it also yields an improvement from a system perspective.
\begin{result}
	If a specific pooling mechanism is profitable for the \gls{abk:msp}, it always yields improvements from a system and a customer perspective by reducing CO\textsubscript{2} emissions, improving the service rate, and reducing travel cost. The benefits for all participants increase with increasing MARs.
\end{result}
From a system perspective, a \gls{abk:ccp} mechanism dominates a \gls{abk:pcp} mechanism. Choosing lower cost parameters $\delta$ and $p^c$ leads to higher distance and emission savings and may foster higher MARs.

From a provider perspective, we assume the mechanism selection to be mostly profit-driven. Accordingly, an \gls{abk:msp} may generally favor a high-cost or medium-cost mechanism, even if this entails a low MAR. This reveals a goal conflict between the system optimal mechanism and the \gls{abk:msp}'s choice. Indeed, a provider prefers high- and medium-cost mechanisms even if they result in vastly lower MARs (cf. Table~\ref{tab:profit_main}).

Our overall results suggest that the analyzed \gls{abk:ccp} mechanisms may often dominate related \gls{abk:pcp} mechanisms. To formally analyze this conjecture, we use the concept of \textit{Pareto dominance}.
\begin{definition}[Pareto Dominance]
	A pooling mechanism $M_1$ (Pareto) dominates another mechanism $M_2$ ($M_1\succ M_2$) if the average profit of the provider in $M_1$ is at least as high as in $M_2$, and the average costs for customers in $M_1$ are at most as high as in $M_2$. If $M_1$ dominates $M_2$ only for a specific MAR range ($\alpha-\beta$), we say that $M_1$ partially dominates $M_2$~($M_1\overset{\alpha-\beta}{\succ}M_2$).
\end{definition}

\begin{figure}[!hb]
	\centering
	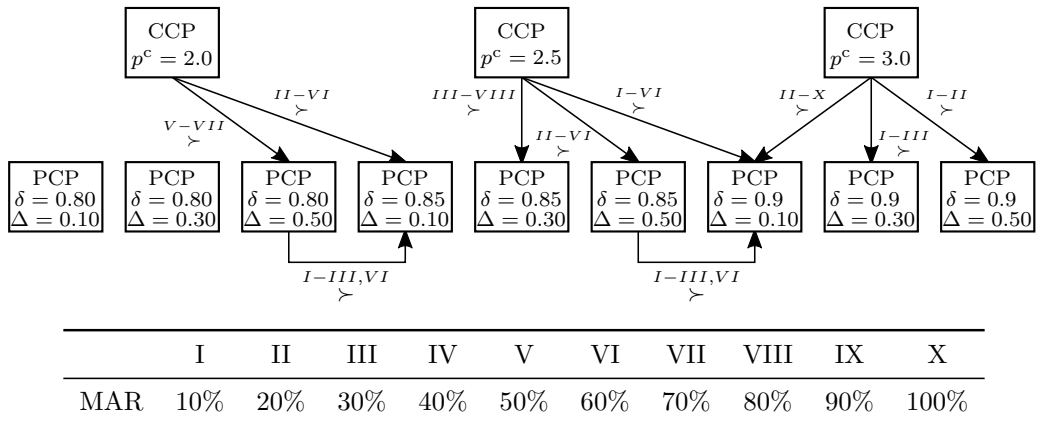
	\singlespacing\footnotesize 
	\begin{tabular}{rcccccccccc}
		\toprule
		& I     & II    & III   & IV    & V     & VI    & VII   & VIII  & IX    & X \\
		\midrule
		MAR   & 10\%  & 20\%  & 30\%  & 40\%  & 50\%  & 60\%  & 70\%  & 80\%  & 90\%  & 100\% \\
		\bottomrule
	\end{tabular}
	\caption{Domination Relationships between mechanisms.\label{fig:dominance}}
	\vspace{-0.4cm}
\end{figure}

Figure \ref{fig:dominance} shows the dominance relationships between the analyzed pooling mechanisms. Here, an arrow and its MAR label visualize the (partial) dominance of a mechanism. These relations allow for the following findings.

Only low-cost PCP mechanisms with $\delta = 0.8$ and $\Delta \leq 0.3$ remain non-dominated because these PCP mechanisms yield the highest customer cost savings. However, these mechanisms yield a maximum provider profit increase of 5.78\% at a MAR of 100\%, which remains below the profit increase most medium and high-cost mechanisms achieve for a MAR of 20\% (see Section~\ref{subsec:providerperspective}).  A CCP counterpart partially dominates all other PCP mechanisms.
\begin{observation}
	Low-cost PCP mechanisms need significantly higher MARs compared to all other mechanisms in order to be profitable for the provider. Still, a CCP mechanism with $p^C = 2.0$ Pareto dominates the most profitable PCP low-cost mechanism for MARs between 50\% and 70\%.
\end{observation}
For medium-cost mechanisms we observe a similar trend.
\begin{observation}
	For medium-cost mechanisms a CCP mechanism with a change fee of $p^C = 2.50$ Pareto dominates PCP mechanisms with $\delta = 0.85, \Delta \ge 0.3$ for MARs between 30\% and 60\%.
\end{observation}

For both the low-cost and the medium-cost mechanisms, the corresponding CCP mechanism leads to higher profits than the \gls{abk:pcp} mechanism for low MARs. For high cost-mechanisms, analogous recommendations remain less obvious. A CCP mechanism clearly dominates a PCP mechanism with $\Delta = 0.1$. For larger $\Delta$, a CCP mechanism only dominates a PCP mechanism up to a MAR of 20\% and 30\%, respectively. For larger MARs, PCP leads to higher profits. 

To synthesize these observations, we note that Figure~\ref{fig:dominance} indicates that for most MARs up to 60\%, both customers and provider would prefer a cheaper PCP mechanism with a $\Delta$ of $0.5$ over a more expensive PCP mechanism with a $\Delta$ of $0.1$. Accordingly, we focus the following synthesis to PCP mechanisms with $\Delta = 0.5$. 

\begin{figure}[!hb]
	\centering
	\input{images/marSensitivity.tikz}
	\caption{Comparison of CCP and PCP mechanisms for different MARs.}
	\label{fig:marSensitivtiy}
	\vspace{-0.7cm}
\end{figure}
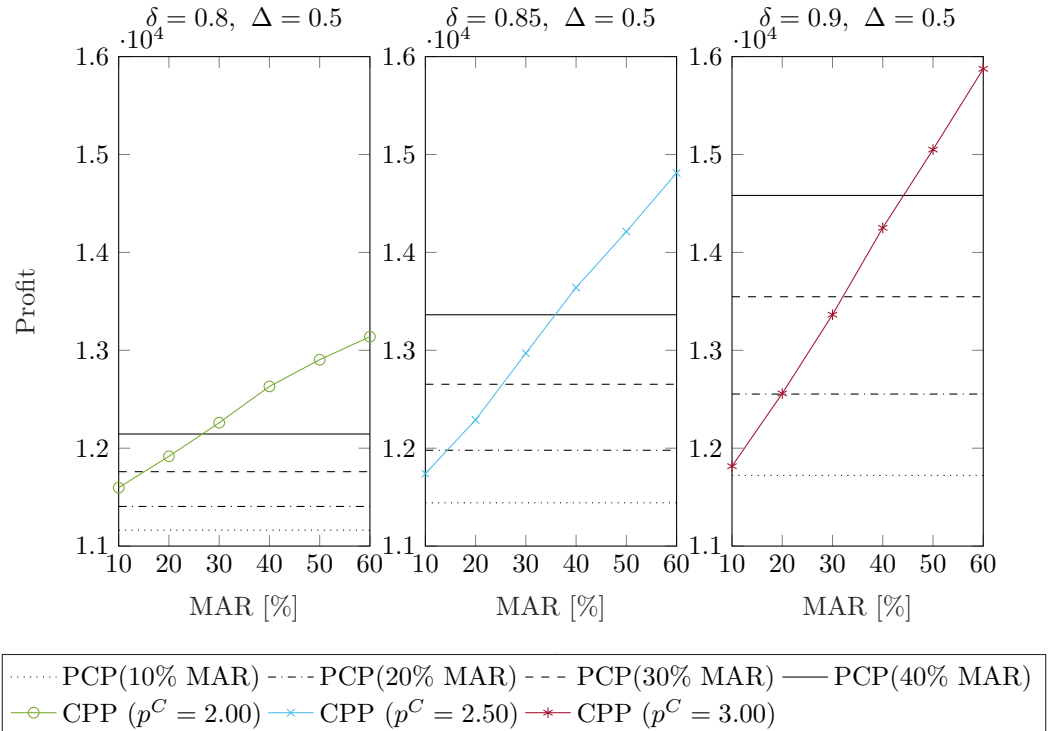

Focusing solely on the analyzed dominance relations points towards the benefit of a CCP mechanism in specific scenarios but does not allow for a general conclusion. While this discussion seems to be fair by definition, we may question the underlying assumption that the CCP and PCP mechanisms will always yield the same MAR share. Accordingly, we relax this assumption in the following and conjecture that a CCP mechanism may yield higher MARs compared to a PCP mechanism, as participating in CCP pooling remains a weekly-dominant strategy for customers. Indeed, a high-cost PCP mechanism might lead to an increase in cost for some customers. Moreover, the fraction of poolable customers that decrease their cost by 10\% or more is significantly larger in a CCP mechanism. 
Figure~\ref{fig:marSensitivtiy} shows the profit for each CCP mechanism and MARs between 10\% and 60\% in relation to the profit 
of the respective PCP mechanism for MARs between 10\% and 40\%. As can be seen, the CCP mechanism yields significantly higher profits in the low-cost scenario, even if the MAR of the CCP would be hypothetically lower than for the PCP setting. For the medium-cost scenario, the CCP mechanism yields higher profits at any MAR than the PCP mechanism, allowing for a significantly increased delta for higher MARs. For the high-cost scenario, the PCP mechanism generates higher profit assuming an equal MAR of 30\% or 40\% compared to the CCP mechanism. However, assuming a slightly higher MAR for the CCP mechanism compared to the PCP mechanism remedies this disadvantage.

In conclusion, a CCP mechanism appears to be the preferable mechanism from a provider profit perspective if it leads to slightly higher MARs than a PCP mechanism. We argue that the lower average cost for customers at high MARs and the better prospect of obtaining noticeable savings without the chance of making a loss (cf. Section \ref{sec:flexPricing}) leads to a higher MAR when employing a CCP mechanism. Moreover, CCP mechanisms always lead to better system performance. Accordingly, our results indicate that operating a ride-pooling system with a CCP mechanism may improve the overall system performance in terms of a higher service rate and less carbon dioxide emissions while creating a win-win situation for fleet operators and customers.

%% file: images/dominance.pdf_tex
%% Creator: Inkscape inkscape 0.92.5, www.inkscape.org
%% Accompanies image file 'dominance.pdf' (pdf, eps, ps)
\begingroup%
  \makeatletter%
  \providecommand\color[2][]{%
    \errmessage{(Inkscape) Color is used for the text in Inkscape, but the package 'color.sty' is not loaded}%
    \renewcommand\color[2][]{}%
  }%
  \providecommand\transparent[1]{%
    \errmessage{(Inkscape) Transparency is used (non-zero) for the text in Inkscape, but the package 'transparent.sty' is not loaded}%
    \renewcommand\transparent[1]{}%
  }%
  \providecommand\rotatebox[2]{#2}%
  \newcommand*\fsize{\dimexpr\f@size pt\relax}%
  \newcommand*\lineheight[1]{\fontsize{\fsize}{#1\fsize}\selectfont}%
  \ifx\svgwidth\undefined%
    \setlength{\unitlength}{384.82299354bp}%
    \ifx\svgscale\undefined%
      \relax%
    \else%
      \setlength{\unitlength}{\unitlength * \real{\svgscale}}%
    \fi%
  \else%
    \setlength{\unitlength}{\svgwidth}%
  \fi%
  \global\let\svgwidth\undefined%
  \global\let\svgscale\undefined%
  \makeatother%
  \begin{picture}(1,0.27436916)%
  \scriptsize
    \lineheight{1}%
    \setlength\tabcolsep{0pt}%
    \put(0,0){\includegraphics[width=\unitlength,page=1]{./images/dominance.pdf}}%
    \put(0.11969716,0.2436031){\color[rgb]{0,0,0}\makebox(0,0)[lt]{\lineheight{1.25}\smash{\begin{tabular}[t]{c}CCP\\$\changeFee=2.0$\end{tabular}}}}%
    \put(0.4582674,0.24443989){\color[rgb]{0,0,0}\makebox(0,0)[lt]{\lineheight{1.25}\smash{\begin{tabular}[t]{c}CCP\\$\changeFee=2.5$\end{tabular}}}}%
    \put(0.79674589,0.24360314){\color[rgb]{0,0,0}\makebox(0,0)[lt]{\lineheight{1.25}\smash{\begin{tabular}[t]{c}CCP\\$\changeFee=3.0$\end{tabular}}}}%
    \put(0.88922037,0.17){\color[rgb]{0,0,0}\makebox(0,0)[lt]{\lineheight{1.25}\smash{\begin{tabular}[t]{l}\tiny$\overset{I-II}{\succ}$\end{tabular}}}}%
    \put(0.84291805,0.13){\color[rgb]{0,0,0}\makebox(0,0)[lt]{\lineheight{1.25}\smash{\begin{tabular}[t]{l}\tiny$\overset{I-III}{\succ}$\end{tabular}}}}%
    \put(0.74628248,0.17){\color[rgb]{0,0,0}\makebox(0,0)[lt]{\lineheight{1.25}\smash{\begin{tabular}[t]{l}\tiny$\overset{II-X}{\succ}$\end{tabular}}}}%
    \put(0.5881417,0.172){\color[rgb]{0,0,0}\makebox(0,0)[lt]{\lineheight{1.25}\smash{\begin{tabular}[t]{l}\tiny$\overset{I-VI}{\succ}$\end{tabular}}}}%
    \put(0.41,0.17){\color[rgb]{0,0,0}\makebox(0,0)[lt]{\lineheight{1.25}\smash{\begin{tabular}[t]{l}\tiny$\overset{III-VIII}{\succ}$\end{tabular}}}}%
    \put(0.51050682,0.13){\color[rgb]{0,0,0}\makebox(0,0)[lt]{\lineheight{1.25}\smash{\begin{tabular}[t]{l}\tiny$\overset{II-VI}{\succ}$\end{tabular}}}}%
    \put(0.25695519,0.17){\color[rgb]{0,0,0}\makebox(0,0)[lt]{\lineheight{1.25}\smash{\begin{tabular}[t]{l}\tiny$\overset{II-VI}{\succ}$\end{tabular}}}}%
    \put(0.15,0.14){\color[rgb]{0,0,0}\makebox(0,0)[lt]{\lineheight{1.25}\smash{\begin{tabular}[t]{l}\tiny$\overset{V-VII}{\succ}$\end{tabular}}}}%
    \put(0.02361926,0.1){\color[rgb]{0,0,0}\makebox(0,0)[lt]{\lineheight{1.25}\smash{\begin{tabular}[t]{l}PCP\end{tabular}}}}%
    \put(0.13599267,0.1){\color[rgb]{0,0,0}\makebox(0,0)[lt]{\lineheight{1.25}\smash{\begin{tabular}[t]{l}PCP\end{tabular}}}}%
    \put(0.24920091,0.1){\color[rgb]{0,0,0}\makebox(0,0)[lt]{\lineheight{1.25}\smash{\begin{tabular}[t]{l}PCP\end{tabular}}}}%
    \put(0.36172522,0.1){\color[rgb]{0,0,0}\makebox(0,0)[lt]{\lineheight{1.25}\smash{\begin{tabular}[t]{l}PCP\end{tabular}}}}%
    \put(0.47531162,0.1){\color[rgb]{0,0,0}\makebox(0,0)[lt]{\lineheight{1.25}\smash{\begin{tabular}[t]{l}PCP\end{tabular}}}}%
    \put(0.58852188,0.1){\color[rgb]{0,0,0}\makebox(0,0)[lt]{\lineheight{1.25}\smash{\begin{tabular}[t]{l}PCP\end{tabular}}}}%
    \put(0.70104619,0.1){\color[rgb]{0,0,0}\makebox(0,0)[lt]{\lineheight{1.25}\smash{\begin{tabular}[t]{l}PCP\end{tabular}}}}%
    \put(0.81402708,0.1){\color[rgb]{0,0,0}\makebox(0,0)[lt]{\lineheight{1.25}\smash{\begin{tabular}[t]{l}PCP\end{tabular}}}}%
    \put(0.92655139,0.1){\color[rgb]{0,0,0}\makebox(0,0)[lt]{\lineheight{1.25}\smash{\begin{tabular}[t]{l}PCP\end{tabular}}}}%
    \put(0.0060923,0.08){\color[rgb]{0,0,0}\makebox(0,0)[lt]{\lineheight{1.25}\smash{\begin{tabular}[t]{l}$\delta = 0.80$\end{tabular}}}}%
    \put(0.11899678,0.08){\color[rgb]{0,0,0}\makebox(0,0)[lt]{\lineheight{1.25}\smash{\begin{tabular}[t]{l}$\delta = 0.80$\end{tabular}}}}%
    \put(0.23190127,0.08){\color[rgb]{0,0,0}\makebox(0,0)[lt]{\lineheight{1.25}\smash{\begin{tabular}[t]{l}$\delta = 0.80$\end{tabular}}}}%
    \put(0.34480578,0.08){\color[rgb]{0,0,0}\makebox(0,0)[lt]{\lineheight{1.25}\smash{\begin{tabular}[t]{l}$\delta = 0.85$\end{tabular}}}}%
    \put(0.45771022,0.08){\color[rgb]{0,0,0}\makebox(0,0)[lt]{\lineheight{1.25}\smash{\begin{tabular}[t]{l}$\delta = 0.85$\end{tabular}}}}%
    \put(0.57061471,0.08){\color[rgb]{0,0,0}\makebox(0,0)[lt]{\lineheight{1.25}\smash{\begin{tabular}[t]{l}$\delta = 0.85$\end{tabular}}}}%
    \put(0.6835192,0.08){\color[rgb]{0,0,0}\makebox(0,0)[lt]{\lineheight{1.25}\smash{\begin{tabular}[t]{l}$\delta = 0.9$\end{tabular}}}}%
    \put(0.79642368,0.08){\color[rgb]{0,0,0}\makebox(0,0)[lt]{\lineheight{1.25}\smash{\begin{tabular}[t]{l}$\delta = 0.9$\end{tabular}}}}%
    \put(0.90932817,0.08){\color[rgb]{0,0,0}\makebox(0,0)[lt]{\lineheight{1.25}\smash{\begin{tabular}[t]{l}$\delta = 0.9$\end{tabular}}}}%
    \put(0.0020923,0.06){\color[rgb]{0,0,0}\makebox(0,0)[lt]{\lineheight{1.25}\smash{\begin{tabular}[t]{l}$\Delta = 0.10$\end{tabular}}}}%
    \put(0.11499678,0.06){\color[rgb]{0,0,0}\makebox(0,0)[lt]{\lineheight{1.25}\smash{\begin{tabular}[t]{l}$\Delta = 0.30$\end{tabular}}}}%
    \put(0.22790127,0.06){\color[rgb]{0,0,0}\makebox(0,0)[lt]{\lineheight{1.25}\smash{\begin{tabular}[t]{l}$\Delta = 0.50$\end{tabular}}}}%
    \put(0.34080578,0.06){\color[rgb]{0,0,0}\makebox(0,0)[lt]{\lineheight{1.25}\smash{\begin{tabular}[t]{l}$\Delta = 0.10$\end{tabular}}}}%
    \put(0.45371022,0.06){\color[rgb]{0,0,0}\makebox(0,0)[lt]{\lineheight{1.25}\smash{\begin{tabular}[t]{l}$\Delta = 0.30$\end{tabular}}}}%
    \put(0.56661471,0.06){\color[rgb]{0,0,0}\makebox(0,0)[lt]{\lineheight{1.25}\smash{\begin{tabular}[t]{l}$\Delta = 0.50$\end{tabular}}}}%
    \put(0.6795192,0.06){\color[rgb]{0,0,0}\makebox(0,0)[lt]{\lineheight{1.25}\smash{\begin{tabular}[t]{l}$\Delta = 0.10$\end{tabular}}}}%
    \put(0.79242368,0.06){\color[rgb]{0,0,0}\makebox(0,0)[lt]{\lineheight{1.25}\smash{\begin{tabular}[t]{l}$\Delta = 0.30$\end{tabular}}}}%
    \put(0.90532817,0.06){\color[rgb]{0,0,0}\makebox(0,0)[lt]{\lineheight{1.25}\smash{\begin{tabular}[t]{l}$\Delta = 0.50$\end{tabular}}}}%
    \put(0.285,-0.012){\color[rgb]{0,0,0}\makebox(0,0)[lt]{\lineheight{1.25}\smash{\begin{tabular}[t]{l}$\overset{I-III,VI}{\succ}$\end{tabular}}}}%
    \put(0.625,-0.012){\color[rgb]{0,0,0}\makebox(0,0)[lt]{\lineheight{1.25}\smash{\begin{tabular}[t]{l}$\overset{I-III,VI}{\succ}$\end{tabular}}}}%
  \end{picture}%
\endgroup%

%% file: images/marSensitivity.tikz
\definecolor{mycolor1}{rgb}{0.46600,0.67400,0.18800}%
\definecolor{mycolor2}{rgb}{0.30100,0.74500,0.93300}%
\definecolor{mycolor3}{rgb}{0.63500,0.07800,0.18400}%
\begin{tikzpicture}
\footnotesize
\begin{axis}[%
width=3.306cm,
height=6.49cm,
at={(0cm,0cm)},
scale only axis,
xmin=1,
xmax=6,
xlabel style={font=\color{white!15!black}},
xlabel={MAR [\%]},
xtick={1,2,3,4,5,6},
xticklabels={10,20,30,40,50,60},
ymin=11000,
ymax=16000,
ylabel style={font=\color{white!15!black}},
ylabel={Profit},
axis background/.style={fill=white},
title style={font=\bfseries},
title={$\delta = 0.8,\ \Delta = 0.5$},
]
\addplot [color=black, dotted]
  table[row sep=crcr]{%
1	11163\\
2	11163\\
3	11163\\
4	11163\\
5	11163\\
6	11163\\
};

\addplot [color=black, dashdotted]
  table[row sep=crcr]{%
1	11405\\
2	11405\\
3	11405\\
4	11405\\
5	11405\\
6	11405\\
};

\addplot [color=black, dashed]
  table[row sep=crcr]{%
1	11760\\
2	11760\\
3	11760\\
4	11760\\
5	11760\\
6	11760\\
};

\addplot [color=black]
  table[row sep=crcr]{%
1	12145\\
2	12145\\
3	12145\\
4	12145\\
5	12145\\
6	12145\\
};

\addplot [color=mycolor1, mark=o, mark options={solid, mycolor1}]
  table[row sep=crcr]{%
1	11597\\
2	11917\\
3	12260\\
4	12631\\
5	12904\\
6	13139\\
};

\end{axis}

\begin{axis}[%
width=3.306cm,
height=6.49cm,
at={(4.035cm,0cm)},
scale only axis,
xmin=1,
xmax=6,
xlabel style={font=\color{white!15!black}},
xlabel={MAR [\%]},
xtick={1,2,3,4,5,6},
xticklabels={10,20,30,40,50,60},
ymin=11000,
ymax=16000,
ylabel style={font=\color{white!15!black}},
axis background/.style={fill=white},
title style={font=\bfseries},
title={$\delta = 0.85,\ \Delta = 0.5$},
]
\addplot [color=black, dotted]
  table[row sep=crcr]{%
1	11442\\
2	11442\\
3	11442\\
4	11442\\
5	11442\\
6	11442\\
};

\addplot [color=black, dashdotted]
  table[row sep=crcr]{%
1	11979\\
2	11979\\
3	11979\\
4	11979\\
5	11979\\
6	11979\\
};

\addplot [color=black, dashed]
  table[row sep=crcr]{%
1	12653\\
2	12653\\
3	12653\\
4	12653\\
5	12653\\
6	12653\\
};

\addplot [color=black]
  table[row sep=crcr]{%
1	13363\\
2	13363\\
3	13363\\
4	13363\\
5	13363\\
6	13363\\
};

\addplot [color=mycolor2, mark=x, mark options={solid, mycolor2}]
  table[row sep=crcr]{%
1	11740\\
2	12290\\
3	12970\\
4	13641\\
5	14212\\
6	14812\\
};

\end{axis}

\begin{axis}[%
width=3.306cm,
height=6.49cm,
at={(8.07cm,0cm)},
scale only axis,
xmin=1,
xmax=6,
xlabel style={font=\color{white!15!black}},
xlabel={MAR [\%]},
xtick={1,2,3,4,5,6},
xticklabels={10,20,30,40,50,60},
ymin=11000,
ymax=16000,
ylabel style={font=\color{white!15!black}},
axis background/.style={fill=white},
title style={font=\bfseries},
title={$\delta = 0.9,\ \Delta = 0.5$},
]
\addplot [color=black, dotted]
  table[row sep=crcr]{%
1	11722\\
2	11722\\
3	11722\\
4	11722\\
5	11722\\
6	11722\\
};

\addplot [color=black, dashdotted]
  table[row sep=crcr]{%
1	12553\\
2	12553\\
3	12553\\
4	12553\\
5	12553\\
6	12553\\
};

\addplot [color=black, dashed]
  table[row sep=crcr]{%
1	13547\\
2	13547\\
3	13547\\
4	13547\\
5	13547\\
6	13547\\
};

\addplot [color=black]
  table[row sep=crcr]{%
1	14582\\
2	14582\\
3	14582\\
4	14582\\
5	14582\\
6	14582\\
};

\addplot [color=mycolor3, mark=asterisk, mark options={solid, mycolor3}]
  table[row sep=crcr]{%
1	11816\\
2	12559\\
3	13364\\
4	14251\\
5	15050\\
6	15875\\
};

\end{axis}

\begin{axis}[%
width=3.306cm,
height=1.49cm,
at={(4.105cm,-2.25cm)},
scale only axis,
xmin=-1,
xmax=1,
ymin=-1,
ymax=1,
axis line style={draw=none},
ticks=none,
legend columns = 4,
legend style={at={(2.45,0.55)}, legend cell align=left, align=left, draw=white!15!black}
]
\addplot [color=black, dotted]
  table[row sep=crcr]{%
0	0\\
};
\addlegendentry{PCP(10\% MAR)}

\addplot [color=black, dashdotted]
  table[row sep=crcr]{%
0	0\\
};
\addlegendentry{PCP(20\% MAR)}

\addplot [color=black, dashed]
  table[row sep=crcr]{%
0	0\\
};
\addlegendentry{PCP(30\% MAR)}

\addplot [color=black]
  table[row sep=crcr]{%
0	0\\
};
\addlegendentry{PCP(40\% MAR)}

\addplot [color=mycolor1, mark=o, mark options={solid, mycolor1}]
  table[row sep=crcr]{%
0	0\\
};
\addlegendentry{CPP ($p^C = 2.00$)}

\addplot [color=mycolor2, mark=x, mark options={solid, mycolor2}]
  table[row sep=crcr]{%
0	0\\
};
\addlegendentry{CPP ($p^C = 2.50$)}

\addplot [color=mycolor3, mark=asterisk, mark options={solid, mycolor3}]
  table[row sep=crcr]{%
0	0\\
};
\addlegendentry{CPP ($p^C = 3.00$)}

\end{axis}
\end{tikzpicture}%

%% file: chapters/conclusion.tex
\section{Conclusion}\label{sec:conclusion}
We studied pooling mechanisms for on-demand pooling in today's ride-hailing fleets. To this end, we presented a customer-centered mechanism (\gls{abk:ccp}) that accounts or each customer's inconvenience and splits resulting cost savings accordingly. We compared this mechanism with a standard provider-centered mechanism (\gls{abk:pcp}) in which the operator pools customers in exchange for a fixed discount. We formally analyzed both mechanisms and showed that it is individually rational for a customer to participate in CCP. Moreover, we showed that this is not the case for PCP. Indeed, PCP may even lead to missed profit opportunities and is most beneficial to unmatched customers. 

We implemented a simulation environment to numerically analyze the different pooling mechanisms for a real-world case study. Our results show that pooling can be beneficial from a system, provider, and customer perspective. We show that ride-pooling may simultaneously lead to a win-win situation for operators and customers while significantly reducing fleet emissions. While the PCP mechanism remains very sensitive to its parameter settings, a CCP mechanism reveals a more stable behavior. In general, a CCP mechanism (partially) Pareto dominates its PCP counterpart. Further, assuming that a CCP mechanism yields a slightly higher percentage of customers that participate in pooling, a CCP mechanism is always preferable for the fleet operator from a profit maximization perspective.

The benefits of CCP mechanisms discussed in this paper open the field or further research. As natural next steps, one may focus on efficient algorithmic implementations to make the algorithm usable in practice, empirical field experiments to verify the impact of a CCP mechanism, or methodologically linking the proposed mechanism to general surge pricing strategies.

%% file: chapters/appendix.tex
%--
\section{Proofs}\label{app:proofs}
\begin{proofU}[Proof of Theorem~\ref{the:individualrational}.]
Let $i$ be a customer. Let $\costSol{i}$ and $\costSol{j}$ be the total cost of customers $i,j$ when they are not pooled. Due to Equation \eqref{eq:profPool}, the two customers are pooled if and only if the total cost when pooling, $\costccp{i,j}$ is lower than $\costSol{i}+\costSol{j}$. Further, the CC pooling mechanism always assigns a payment to $i$ such that the resulting costs $\cost{i} < \costSol{i}$. Hence, customers never regret being pooled. On the other hand, whenever a poolable customer takes a solitary ride, she does not incur any additional costs compared to the scenario where she did not set herself poolable. Hence, setting oneself poolable is individually rational. 
\hfill$\Box$
\end{proofU}

\begin{proofU}[Proof of Corollary~\ref{cor:weaklyDominant}.]
Consider customer $i$. The \gls{abk:msp} is bound by the CC pooling mechanism to only pool $i$ when her individual total cost decreases. No action of other customers (e.g., lying about their attributes) can lead to a situation where setting herself not poolable would lead to higher individual costs for $i$ than when setting herself poolable. Therefore, setting herself poolable is a weakly dominant strategy.
\hfill$\Box$
\end{proofU}

\begin{proofU}[Proof of Theorem~\ref{the:notweakDom}]
Let $i$ be a customer with value of time $\valueofTime_i$ that requests a trip from $\origin_i$ to $\destination_i$. Taking a solitary ride, the MSP would charge a price of $\paysol{i} = \fixedPrice + \variablePrice{\origin_i}{\destination_i}$ with arrival time at $t^{\text{d,S}}_i$. Hence, the total cost of $i$ for a solitary ride sum up to $\costSol{i} = \paysol{i} + \valueofTime_i \cdot \left( t^{\text{d,S}}_i  - \pickupTime_i \right)$. Given a discount factor $\discount$ and a detour factor $\deviation$, the total cost of $i$ when pooling can increase up to $\widehat{\costpcp{i}} = \discount\cdot\paysol{i} + (1+\deviation)\cdot \valueofTime_i \cdot \left( t^{\text{d,S}}_i  - \pickupTime_i \right)$. Setting herself poolable ceases to be individual rational when 
\[
\valueofTime_i > \frac{(1-\discount)\cdot\paysol{i}}{\deviation\cdot (t^{\text{d,S}}_i-\pickupTime_i)}.
\]
Hence, customers with high values of time might be better of not setting themselves poolable.
\hfill$\Box$
\end{proofU}

\begin{proofU}[Proof of Theorem~\ref{the:uprofitable}.]
Consider two poolable customers $i$ and $j$ with the same value of time. Assume a PC pooling mechanism with a fixed discount factor of $\discount$ and an arrival time guarantee such that the duration of $i$ and $j$'s trips when pooled is at most by a factor $(1+\deviation)$ longer than if they would travel alone. 

Let customer $i$ requests a ride from location $l_{i}$ to location $l_{D}$. At the same time, customer $j$ requests a ride from $l_{j}$ to $l_{D}$ (thus they both share the same destination location). Let $dist_{ab}$ denote the distance between locations $l_a$ and $l_b$ and let $t_{ab}$ be the travel time between the respective locations. Assume that there are two vehicles, one at $l_i$ and one at $l_j$, which both could pick up $i$ and $j$ immediately. Let travel times and distances be such that $dist_{iD} = dist_{jD}$, $t_{iD} = t_{jD}$, and $t_{ij} = \deviation \cdot t_{iD}$. Hence, $t_{ij} + t_{jD} = (1+\deviation) t_{iD} = (1+\deviation) t_{jD}$, i.e., pooling $i$ and $j$ is allowed. Further, let distances be such that $dist_{ij} < dist_{iD}$ (i.e., pooling $i$ and $j$ is profitable for the MSP since both customers receive a discount regardless of pooling and hence maximizing the MSPs profit is equivalent to minimizing the total driving distance). 

Slightly alter $i$'s and $j$'s origin location to $l_{\tilde{i}}$ and $l_{\tilde{j}}$, such that $t_{\tilde{i}\tilde{j}} = t_{ij}$ and $dist_{\tilde{i}\tilde{j}} = dist_{ij}$, but $t_{\tilde{j}D} = t_{jD} - \epsilon_t = t_{iD} - \epsilon_t$ for some small $\epsilon_t > 0$, $dist_{\tilde{j}D} = dist_{jD} - \epsilon_d$ for some small $\epsilon_d > 0$, and $dist_{\tilde{i}D} = dist_{iD} + \epsilon_d$ as well as $t_{\tilde{i}D} = t_{iD} + \epsilon_t$. Figure \ref{fig:unprofitable} depicts these two scenarios.
\begin{figure}[!ht]
	\begin{center}
		\includegraphics[width=0.5\columnwidth]{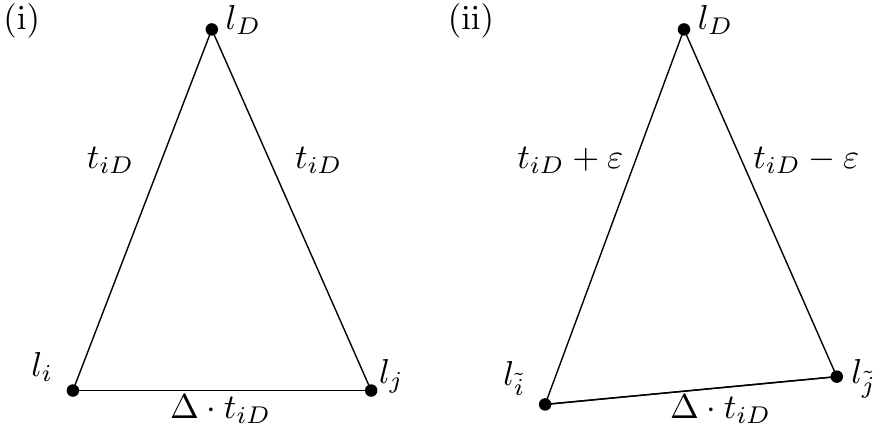}%
		\caption{Example where the PC mechanism fails: While both customers would be pooled in (i), they would not be pooled in the (ii). If there was a profitable pooling opportunity in (ii), customers would not be pooled. If pooling is not profitable in (ii) it would also not be profitable in (i), even though customers are pooled within the PC mechanism.}
		\label{fig:unprofitable}
	\end{center}
\end{figure} 

In the following, we show that either pooling is profitable for all participants with the original and the altered locations, or it is unprofitable for customers in both scenarios.

Assume first that pooling with the altered origin destination would be profitable for both customers in a CC pooling mechanism as compared to taking solitary rides. Pooling is no longer allowed in PC due to the arrival time guarantee since $t_{\tilde{i}\tilde{j}} + t_{\tilde{j}D} > (1+\deviation)t_{\tilde{j}D}$, even though it is profitable for customers and the MSP due to $dist_{\tilde{i}\tilde{j}} + dist_{\tilde{j}D} < dist_{ij} + dist_{jD}$ and $dist_{\tilde{i}D} + dist_{\tilde{j}D} = dist_{iD} + dist_{jD}$. Therefore, the PC mechanism misses a profitable pooling opportunity in this case.

Now, assume that the pooling with the altered origin and destination would not be profitable for both customers in any CC or PC pooling mechanism. Then, it can also not be profitable with the initial origin locations. The sum of total costs for solitary rides is the same in both instances. However, the pooled ride's total cost is lower with the altered locations due to lower distance and time. Hence, if pooling is not profitable for customers in the altered instance, it cannot be profitable in the original instances. The customers are pooled in the original instance's PCP mechanism, even though it is unprofitable for them.
\hfill$\Box$
\end{proofU}

%% file: karaenkeSchifferWaldherr.bbl
\begin{thebibliography}{36}
\expandafter\ifx\csname natexlab\endcsname\relax\def\natexlab#1{#1}\fi
\providecommand{\url}[1]{\texttt{#1}}
\providecommand{\href}[2]{#2}
\providecommand{\path}[1]{#1}
\providecommand{\DOIprefix}{doi:}
\providecommand{\ArXivprefix}{arXiv:}
\providecommand{\URLprefix}{URL: }
\providecommand{\Pubmedprefix}{pmid:}
\providecommand{\doi}[1]{\href{http://dx.doi.org/#1}{\path{#1}}}
\providecommand{\Pubmed}[1]{\href{pmid:#1}{\path{#1}}}
\providecommand{\bibinfo}[2]{#2}
\ifx\xfnm\relax \def\xfnm[#1]{\unskip,\space#1}\fi
%Type = Article
\bibitem[{Agatz et~al.(2012)Agatz, Erera, Savelsbergh \&
  Wang}]{AgatzEreraEtAl2012}
\bibinfo{author}{Agatz, N.}, \bibinfo{author}{Erera, A.},
  \bibinfo{author}{Savelsbergh, M.}, \& \bibinfo{author}{Wang, X.}
  (\bibinfo{year}{2012}).
\newblock \bibinfo{title}{Optimization for dynamic ride-sharing: A review}.
\newblock {\it \bibinfo{journal}{European Journal of Operational Research}\/},
  {\it \bibinfo{volume}{223}\/}, \bibinfo{pages}{295 -- 303}.
%Type = Article
\bibitem[{Agussurja et~al.(2019)Agussurja, Cheng \&
  Lau}]{AgussurjaChengEtAl2019}
\bibinfo{author}{Agussurja, L.}, \bibinfo{author}{Cheng, S.-F.}, \&
  \bibinfo{author}{Lau, H.~C.} (\bibinfo{year}{2019}).
\newblock \bibinfo{title}{A stage aggregation approach for stochastic
  multiperiod last-mile ride-sharing problems}.
\newblock {\it \bibinfo{journal}{Transportation Science}\/}, .
%Type = Article
\bibitem[{Alonso-Mora et~al.(2017)Alonso-Mora, Samaranayake, Wallar, Frazzoli
  \& Rus}]{Alonso-MoraSamaranayakeEtAl2017}
\bibinfo{author}{Alonso-Mora, J.}, \bibinfo{author}{Samaranayake, S.},
  \bibinfo{author}{Wallar, A.}, \bibinfo{author}{Frazzoli, E.}, \&
  \bibinfo{author}{Rus, D.} (\bibinfo{year}{2017}).
\newblock \bibinfo{title}{On-demand high-capacity ride-sharing via dynamic
  trip-vehicle assignment}.
\newblock {\it \bibinfo{journal}{Proceedings of the National Academy of
  Sciences}\/},  {\it \bibinfo{volume}{3}\/}, \bibinfo{pages}{462--467}.
%Type = Misc
\bibitem[{Baltic et~al.(2019)Baltic, Cappy, Hensley \&
  Pfaff}]{BalticCappyEtAl2019}
\bibinfo{author}{Baltic, T.}, \bibinfo{author}{Cappy, A.},
  \bibinfo{author}{Hensley, R.}, \& \bibinfo{author}{Pfaff, N.}
  (\bibinfo{year}{2019}).
\newblock \bibinfo{title}{How sharing the road is likely to transform american
  mobility}.
\newblock \bibinfo{note}{Url:
  \url{https://www.mckinsey.com/industries/automotive-and-assembly/our-insights/how-sharing-the-road-is-likely-to-transform-american-mobility}
  (last accessed: November 6\textsuperscript{th} 2020)}.
%Type = Article
\bibitem[{Berbeglia et~al.(2010)Berbeglia, Cordeau \&
  Laporte}]{BerbigliaCordeauEtAl2010}
\bibinfo{author}{Berbeglia, G.}, \bibinfo{author}{Cordeau, J.-F.}, \&
  \bibinfo{author}{Laporte, G.} (\bibinfo{year}{2010}).
\newblock \bibinfo{title}{Dynamic pickup and delivery problems}.
\newblock {\it \bibinfo{journal}{European Journal of Operational Research}\/},
  {\it \bibinfo{volume}{202}\/}, \bibinfo{pages}{8--15}.
%Type = Article
\bibitem[{Bian \& Liu(2019)}]{BianLiu2019}
\bibinfo{author}{Bian, Z.}, \& \bibinfo{author}{Liu, X.}
  (\bibinfo{year}{2019}).
\newblock \bibinfo{title}{Mechanism design for first-mile ridesharing based on
  personalized requirements part i: Theoretic alanalysis in generalized
  scenarios}.
\newblock {\it \bibinfo{journal}{Transportation Research Part B:
  Methodological}\/},  {\it \bibinfo{volume}{120}\/},
  \bibinfo{pages}{147--171}.
%Type = Article
\bibitem[{Bimpikis et~al.(2019)Bimpikis, Candogan \&
  Saban}]{BimpikisCandoganEtAl2019}
\bibinfo{author}{Bimpikis, K.}, \bibinfo{author}{Candogan, O.}, \&
  \bibinfo{author}{Saban, D.} (\bibinfo{year}{2019}).
\newblock \bibinfo{title}{Spatial pricing in ride-sharing networks}.
\newblock {\it \bibinfo{journal}{Operations Research}\/},  {\it
  \bibinfo{volume}{67}\/}, \bibinfo{pages}{744--769}.
%Type = Article
\bibitem[{B{\"o}sch et~al.(2018)B{\"o}sch, Becker, Becker \&
  Axhausen}]{BoeschBeckerEtAl2018}
\bibinfo{author}{B{\"o}sch, P.~M.}, \bibinfo{author}{Becker, F.},
  \bibinfo{author}{Becker, H.}, \& \bibinfo{author}{Axhausen, K.~W.}
  (\bibinfo{year}{2018}).
\newblock \bibinfo{title}{Cost-based analysis of autonomous mobility services}.
\newblock {\it \bibinfo{journal}{Transport Policy}\/},  {\it
  \bibinfo{volume}{64}\/}, \bibinfo{pages}{76--91}.
%Type = Article
\bibitem[{CBI(2020)}]{CBI2020}
\bibinfo{author}{CBI} (\bibinfo{year}{2020}).
\newblock \bibinfo{title}{How {U}ber makes money now}.
\newblock {\it \bibinfo{journal}{CB Insights Report}\/}, .
\newblock \bibinfo{note}{Available at:
  \url{https://www.cbinsights.com/research/report/how-uber-makes-money/}}.
%Type = Article
\bibitem[{Chen et~al.(2019)Chen, Mes, Schutten \& Quint}]{ChenMesEtAl2019}
\bibinfo{author}{Chen, W.}, \bibinfo{author}{Mes, M.},
  \bibinfo{author}{Schutten, M.}, \& \bibinfo{author}{Quint, J.}
  (\bibinfo{year}{2019}).
\newblock \bibinfo{title}{A ride-sharing problem with meeting points and return
  restrictions}.
\newblock {\it \bibinfo{journal}{Transportation Science}\/},  (pp.
  \bibinfo{pages}{1--26}).
%Type = Article
\bibitem[{Chen \& Wang(2018)}]{ChenWang2018}
\bibinfo{author}{Chen, Y.}, \& \bibinfo{author}{Wang, H.}
  (\bibinfo{year}{2018}).
\newblock \bibinfo{title}{Pricing for a last-mile transportation system}.
\newblock {\it \bibinfo{journal}{Transportation Research Part B:
  Methodological}\/},  {\it \bibinfo{volume}{107}\/}, \bibinfo{pages}{57--69}.
%Type = Article
\bibitem[{Feng et~al.(2020)Feng, Kong \& Wang}]{FengEtAl2020}
\bibinfo{author}{Feng, G.}, \bibinfo{author}{Kong, G.}, \&
  \bibinfo{author}{Wang, Z.} (\bibinfo{year}{2020}).
\newblock \bibinfo{title}{We are on the way: Analysis of on-demand ride-hailing
  systems}.
\newblock {\it \bibinfo{journal}{Manufacturing \& Service Operations
  Management}\/}, .
%Type = Article
\bibitem[{Furuhata et~al.(2013)Furuhata, Dessouky, Ord{\'o}{\~n}ez, Brunet,
  Wang \& Koenig}]{FuruhataDessoukyEtAl2013}
\bibinfo{author}{Furuhata, M.}, \bibinfo{author}{Dessouky, M.},
  \bibinfo{author}{Ord{\'o}{\~n}ez, F.}, \bibinfo{author}{Brunet, M.-E.},
  \bibinfo{author}{Wang, X.}, \& \bibinfo{author}{Koenig, S.}
  (\bibinfo{year}{2013}).
\newblock \bibinfo{title}{Ridesharing: The state-of-the-art and future
  directions}.
\newblock {\it \bibinfo{journal}{Transportation Research Part B:
  Methodological}\/},  {\it \bibinfo{volume}{57}\/}, \bibinfo{pages}{28--46}.
%Type = Article
\bibitem[{Ho et~al.(2018)Ho, Szeto, Kuo, Leung, Petering \&
  Tou}]{HoSzetoEtAl2018}
\bibinfo{author}{Ho, S.~C.}, \bibinfo{author}{Szeto, W.}, \bibinfo{author}{Kuo,
  Y.-H.}, \bibinfo{author}{Leung, J.~M.}, \bibinfo{author}{Petering, M.}, \&
  \bibinfo{author}{Tou, T.~W.} (\bibinfo{year}{2018}).
\newblock \bibinfo{title}{A survey of dial-a-ride problems: Literature review
  and recent developments}.
\newblock {\it \bibinfo{journal}{Transportation Research Part B:
  Methodological}\/},  {\it \bibinfo{volume}{111}\/},
  \bibinfo{pages}{395--421}.
%Type = Article
\bibitem[{Hosni et~al.(2014)Hosni, Naoum-Sawaya \& Artail}]{HosniEtAl2014}
\bibinfo{author}{Hosni, H.}, \bibinfo{author}{Naoum-Sawaya, J.}, \&
  \bibinfo{author}{Artail, H.} (\bibinfo{year}{2014}).
\newblock \bibinfo{title}{The shared-taxi problem: Formulation and solution
  methods}.
\newblock {\it \bibinfo{journal}{Transportation Research Part B:
  Methodological}\/},  {\it \bibinfo{volume}{70}\/}, \bibinfo{pages}{303--318}.
%Type = Article
\bibitem[{Hu(2017)}]{Hu2017}
\bibinfo{author}{Hu, W.} (\bibinfo{year}{2017}).
\newblock \bibinfo{title}{Your uber car creates congestion. should you pay a
  feeto ride?}
\newblock {\it \bibinfo{journal}{Ne York Times}\/},  {\it
  \bibinfo{volume}{available online}\/}.
%Type = Article
\bibitem[{Ke et~al.(2020)Ke, Yang, Li, Wang \& Ye}]{KeEtAl2020}
\bibinfo{author}{Ke, J.}, \bibinfo{author}{Yang, H.}, \bibinfo{author}{Li, X.},
  \bibinfo{author}{Wang, H.}, \& \bibinfo{author}{Ye, J.}
  (\bibinfo{year}{2020}).
\newblock \bibinfo{title}{Pricing and equilibrium in on-demand ride-pooling
  markets}.
\newblock {\it \bibinfo{journal}{Transportation Research Part B:
  Methodological}\/},  {\it \bibinfo{volume}{139}\/},
  \bibinfo{pages}{411--431}.
%Type = Inproceedings
\bibitem[{Lanzetti et~al.(2020)Lanzetti, Schiffer, Ostrovsky \&
  Pavone}]{LanzettiSchifferEtAl2020}
\bibinfo{author}{Lanzetti, N.}, \bibinfo{author}{Schiffer, M.},
  \bibinfo{author}{Ostrovsky, M.}, \& \bibinfo{author}{Pavone, M.}
  (\bibinfo{year}{2020}).
\newblock \bibinfo{title}{On the interplay between self-driving cars and public
  transportation: A game-theoretic perspective}.
\newblock In {\it \bibinfo{booktitle}{Proceedings of the TSL Second Triennial
  Conference}\/}.
%Type = Article
\bibitem[{Long et~al.(2018)Long, Tan, Szeto \& Li}]{LongTanEtAl2018}
\bibinfo{author}{Long, J.}, \bibinfo{author}{Tan, W.}, \bibinfo{author}{Szeto,
  W.}, \& \bibinfo{author}{Li, Y.} (\bibinfo{year}{2018}).
\newblock \bibinfo{title}{Ride-sharing with travel time uncertainty}.
\newblock {\it \bibinfo{journal}{Transportation Research Part B:
  Methodological}\/},  {\it \bibinfo{volume}{118}\/},
  \bibinfo{pages}{143--171}.
%Type = Article
\bibitem[{Lu \& Quadrifoglio(2019)}]{LuQuadrifoglio2019}
\bibinfo{author}{Lu, W.}, \& \bibinfo{author}{Quadrifoglio, L.}
  (\bibinfo{year}{2019}).
\newblock \bibinfo{title}{Fair cost allocation for ridesharing services –
  modeling, mathematical programming and an algorithm to find the nucleolus}.
\newblock {\it \bibinfo{journal}{Transportation Research Part B:
  Methodological}\/},  {\it \bibinfo{volume}{121}\/}, \bibinfo{pages}{41--55}.
%Type = Article
\bibitem[{Morris et~al.(2020)Morris, Zhou, Brown, Khan, Derochers, Campbell,
  Pratt \& Chowdhury}]{MorrisEtAl2020}
\bibinfo{author}{Morris, E.~A.}, \bibinfo{author}{Zhou, Y.},
  \bibinfo{author}{Brown, A.}, \bibinfo{author}{Khan, S.},
  \bibinfo{author}{Derochers, J.~L.}, \bibinfo{author}{Campbell, H.},
  \bibinfo{author}{Pratt, A.~N.}, \& \bibinfo{author}{Chowdhury, M.}
  (\bibinfo{year}{2020}).
\newblock \bibinfo{title}{Are drivers cool with pool? driver attitudes towards
  the shared tnc services uberpool and lyft shared}.
\newblock {\it \bibinfo{journal}{Transport Policy}\/}, .
%Type = Article
\bibitem[{Peng et~al.(2020)Peng, Shan, Jia, Yu, Jiang \&
  Yao}]{PengShanEtAl2020}
\bibinfo{author}{Peng, Z.}, \bibinfo{author}{Shan, W.}, \bibinfo{author}{Jia,
  P.}, \bibinfo{author}{Yu, B.}, \bibinfo{author}{Jiang, Y.}, \&
  \bibinfo{author}{Yao, B.} (\bibinfo{year}{2020}).
\newblock \bibinfo{title}{Stable ride‑sharing matching for the commuters with
  payment design}.
\newblock {\it \bibinfo{journal}{Transportation}\/},  {\it
  \bibinfo{volume}{47}\/}, \bibinfo{pages}{1--21}.
%Type = Article
\bibitem[{Pratt et~al.(2019)Pratt, Morris, Zhou, Khan \&
  Chowdhury}]{PrattEtAl2019}
\bibinfo{author}{Pratt, A.~N.}, \bibinfo{author}{Morris, E.~A.},
  \bibinfo{author}{Zhou, Y.}, \bibinfo{author}{Khan, S.}, \&
  \bibinfo{author}{Chowdhury, M.} (\bibinfo{year}{2019}).
\newblock \bibinfo{title}{What do riders tweet about the people that they meet?
  analyzing online commentary about uberpool and lyft shared/lyft line}.
\newblock {\it \bibinfo{journal}{Transportation research part F: traffic
  psychology and behaviour}\/},  {\it \bibinfo{volume}{62}\/},
  \bibinfo{pages}{459--472}.
%Type = Article
\bibitem[{Qi et~al.(2018)Qi, Li, Liu \& Shen}]{QiLiEtAl2018}
\bibinfo{author}{Qi, W.}, \bibinfo{author}{Li, L.}, \bibinfo{author}{Liu, S.},
  \& \bibinfo{author}{Shen, Z.-J.~M.} (\bibinfo{year}{2018}).
\newblock \bibinfo{title}{Shared mobility for last mile delivery: Design,
  operational prescriptions and environmental impact}.
\newblock {\it \bibinfo{journal}{Manufacturing \& Service Operations
  Management}\/},  {\it \bibinfo{volume}{20}\/}, \bibinfo{pages}{737--751}.
%Type = Article
\bibitem[{Qian et~al.(2017)Qian, Zhang, Ukkusuri \& Yang}]{QianZhangEtAl2017}
\bibinfo{author}{Qian, X.}, \bibinfo{author}{Zhang, W.},
  \bibinfo{author}{Ukkusuri, S.}, \& \bibinfo{author}{Yang, C.}
  (\bibinfo{year}{2017}).
\newblock \bibinfo{title}{Optimal assignment and incentive design in the taxi
  group ride problem}.
\newblock {\it \bibinfo{journal}{Transportation Research Part B:
  Methodological}\/},  {\it \bibinfo{volume}{103}\/}.
%Type = Article
\bibitem[{Rasulkhani \& Chow(2019)}]{RasulkhaniChow2019}
\bibinfo{author}{Rasulkhani, S.}, \& \bibinfo{author}{Chow, J.}
  (\bibinfo{year}{2019}).
\newblock \bibinfo{title}{Route-cost-assignment with joint user and operator
  behavior as a many-to-one stable matching assignment game}.
\newblock {\it \bibinfo{journal}{Transportation Research Part B:
  Methodological}\/},  {\it \bibinfo{volume}{124}\/}, \bibinfo{pages}{60--81}.
%Type = Article
\bibitem[{Shapley(1953)}]{Shapley1953}
\bibinfo{author}{Shapley, L.~S.} (\bibinfo{year}{1953}).
\newblock \bibinfo{title}{A value for n-person games}.
\newblock {\it \bibinfo{journal}{Contributions to the Theory of Games}\/},
  {\it \bibinfo{volume}{2}\/}, \bibinfo{pages}{307--317}.
%Type = Article
\bibitem[{Vazifeh et~al.(2018)Vazifeh, Santi, Resta, Strogatz \&
  Ratti}]{VazifehSantiEtAl2018}
\bibinfo{author}{Vazifeh, M.~M.}, \bibinfo{author}{Santi, P.},
  \bibinfo{author}{Resta, G.}, \bibinfo{author}{Strogatz, S.~H.}, \&
  \bibinfo{author}{Ratti, C.} (\bibinfo{year}{2018}).
\newblock \bibinfo{title}{Addressing the minimum fleet problem in on-demand
  urban mobility}.
\newblock {\it \bibinfo{journal}{Nature}\/},  {\it \bibinfo{volume}{557}\/},
  \bibinfo{pages}{534--538}.
%Type = Article
\bibitem[{Wadud(2017)}]{Wadud2017}
\bibinfo{author}{Wadud, Z.} (\bibinfo{year}{2017}).
\newblock \bibinfo{title}{Fully automated vehicles: A cost of ownership
  analysis to inform early adoption}.
\newblock {\it \bibinfo{journal}{Transportation Research Part A: Policy and
  Practice}\/},  {\it \bibinfo{volume}{101}\/}, \bibinfo{pages}{163--176}.
%Type = Article
\bibitem[{Wang(2019)}]{Wang2019}
\bibinfo{author}{Wang, H.} (\bibinfo{year}{2019}).
\newblock \bibinfo{title}{Routing and scheduling for a last-mile transportation
  system}.
\newblock {\it \bibinfo{journal}{Transportation Science}\/},  {\it
  \bibinfo{volume}{53}\/}, \bibinfo{pages}{131--147}.
%Type = Article
\bibitem[{Wang \& Yang(2019)}]{WangYang2019}
\bibinfo{author}{Wang, H.}, \& \bibinfo{author}{Yang, H.}
  (\bibinfo{year}{2019}).
\newblock \bibinfo{title}{Ridesourcing systems: A framework and review}.
\newblock {\it \bibinfo{journal}{Transportation Research Part B:
  Methodological}\/},  {\it \bibinfo{volume}{129}\/},
  \bibinfo{pages}{122--155}.
%Type = Article
\bibitem[{Wang et~al.(2017)Wang, Agatz \& Erera}]{WangAgatzEtAl2017}
\bibinfo{author}{Wang, X.}, \bibinfo{author}{Agatz, N.}, \&
  \bibinfo{author}{Erera, A.} (\bibinfo{year}{2017}).
\newblock \bibinfo{title}{Stable matching for dynamic ride-sharing systems}.
\newblock {\it \bibinfo{journal}{Transportation Science}\/},  {\it
  \bibinfo{volume}{52}\/}, \bibinfo{pages}{850--867}.
%Type = Article
\bibitem[{Wang et~al.(2018)Wang, Yang \& Zhu}]{WangYangEtAl2018}
\bibinfo{author}{Wang, X.}, \bibinfo{author}{Yang, H.}, \&
  \bibinfo{author}{Zhu, D.} (\bibinfo{year}{2018}).
\newblock \bibinfo{title}{Driver-rider cost-sharing strategies and equilibria
  in a ridesharing program}.
\newblock {\it \bibinfo{journal}{Transportation Science}\/},  {\it
  \bibinfo{volume}{52}\/}, \bibinfo{pages}{868--881}.
%Type = Article
\bibitem[{Yan et~al.(2020)Yan, Zhu, Korolko \& Woodard}]{YanZhuEtAl2020}
\bibinfo{author}{Yan, C.}, \bibinfo{author}{Zhu, H.}, \bibinfo{author}{Korolko,
  N.}, \& \bibinfo{author}{Woodard, D.} (\bibinfo{year}{2020}).
\newblock \bibinfo{title}{Dynamic pricing and matching in ride-hailing
  platforms}.
\newblock {\it \bibinfo{journal}{Naval Research Logistics}\/},  {\it
  \bibinfo{volume}{67}\/}, \bibinfo{pages}{705--724}.
%Type = Article
\bibitem[{Yang et~al.(2020{\natexlab{a}})Yang, Ren, Legara, Li, Ong, Lin \&
  Monterola}]{YangRenEtAl2020}
\bibinfo{author}{Yang, B.}, \bibinfo{author}{Ren, S.}, \bibinfo{author}{Legara,
  E.}, \bibinfo{author}{Li, Z.}, \bibinfo{author}{Ong, E.},
  \bibinfo{author}{Lin, L.}, \& \bibinfo{author}{Monterola, C.}
  (\bibinfo{year}{2020}{\natexlab{a}}).
\newblock \bibinfo{title}{Phase transition in taxi dynamics and impact of
  ridesharing}.
\newblock {\it \bibinfo{journal}{Transportation Science}\/},  {\it
  \bibinfo{volume}{54}\/}.
%Type = Article
\bibitem[{Yang et~al.(2020{\natexlab{b}})Yang, Qin, Ke \& Ye}]{YangQinEtAl2020}
\bibinfo{author}{Yang, H.}, \bibinfo{author}{Qin, X.}, \bibinfo{author}{Ke,
  J.}, \& \bibinfo{author}{Ye, J.} (\bibinfo{year}{2020}{\natexlab{b}}).
\newblock \bibinfo{title}{Optimizing matching time interval and matching radius
  in on-demand ride-sourcing markets}.
\newblock {\it \bibinfo{journal}{Transportation Research Part B:
  Methodological}\/},  {\it \bibinfo{volume}{131}\/}, \bibinfo{pages}{84--105}.

\end{thebibliography}
